\documentclass[11pt]{article}

\usepackage{amsfonts,amssymb,jheppub,dsfont,graphicx,mathrsfs,tikz,physics,bbm}
\allowdisplaybreaks

\usepackage[numbers,sort&compress]{natbib}

\usepackage[most]{tcolorbox}
\tcbset{colback=yellow!10!white,colframe=black}

\newcommand{\cratio}{x}

\newcommand{\bs}[1]{\boldsymbol{#1}}
\newcommand{\iu}{\mathrm{i}\mkern1mu}
\newcommand{\ju}{\mathrm{j}\mkern1mu}

\newcommand{\Mdu}[3]{{#1}_{#2}^{\phantom{#2}#3}}

\title{Virasoro OPE and Conformal Blocks from the Inverse Shapovalov Form}

\author{Jean-Fran\c{c}ois Fortin$^{\ast}$, Lorenzo Quintavalle$^{\ast,\S,\ddagger}$, and Witold Skiba$^{\dagger}$}

\affiliation{
$^\ast$D\'epartement de Physique, de G\'enie Physique et d'Optique, Universit\'e Laval, Qu\'ebec, QC~G1V~0A6, Canada}
\affiliation{$^\S$ Dipartimento di Fisica, Università di Torino, Via P. Giuria 1, 10125 Torino, Italy}
\affiliation{$^\ddagger$ INFN - Sezione di Torino, Via P. Giuria 1, 10125 Torino, Italy}
\affiliation{$^\dagger$Department of Physics, Yale University, New Haven, CT 06520, USA
}

\emailAdd{jean-francois.fortin@phy.ulaval.ca}
\emailAdd{lorenzo.quintavalle@unito.it}
\emailAdd{witold.skiba@yale.edu}

\abstract{We derive expressions for the Virasoro OPE and four-point conformal blocks on the sphere via the resolution of identity recently determined in~\cite{Fortin:2024xir}.  Even though the resolution of the identity depends on Virasoro singular vectors, our expression for the blocks does not depend on their precise form, but just on their well-known conformal weights.  We verify that our expression is compatible with---but differs from---Zamolodchikov's $h$-recursion relation and we also examine the impact of various large central charge limits in our formula.  A \textit{Mathematica} notebook with a simple implementation of our expression for the Virasoro conformal blocks is provided as an ancillary file.
}

\date{September 2025}

\begin{document}

\maketitle



\section{Introduction}

Conformal field theories (CFTs) in two dimensions form an important subset of quantum field theories, and have been a central research topic for more than forty years~\cite{Belavin:1984vu} (for some recent reviews see~\cite{Ribault:2014hia,Teschner:2017del,Kusuki:2024gtq}).  Their universal nature, exemplified by the special role they play in the renormalization group flow, makes them relevant for multiple fields of physics. These range from two-dimensional critical phenomena~\cite{Cardy:2008jc}, to string theory~\cite{Polchinski:books}, integrable systems~\cite{Bazhanov:1994ft}, holography~\cite{Heemskerk:2009pn}, and much more. Their invariance under an infinite conformal symmetry group, manifested locally via two copies of the Virasoro algebra,\footnote{For our purposes, we will focus on the holomorphic part of the $2d$ CFT conformal algebra since the extension of our results to the two copies (holomorphic and anti-holomorphic) is straightforward.} allows for an improved control of the $2d$-CFT correlators and is central to the wide success of these theories.  The combination of conformal invariance with the convergent Virasoro operator product expansion (OPE) allows to reorganize all the dynamical data of the theories in the spectrum of operators, made of holomorphic and anti-holomorphic conformal weights $(h,\bar{h})$ for each field, and the three-point dynamical coefficients that appear in the OPE.  In order to determine the higher-point correlators from this data, however, one has to combine it with the knowledge of a ``basis of functions'' for Virasoro correlators, the \emph{Virasoro conformal blocks}. These are special functions associated with the OPE decomposition of a correlator which repackage the contribution of each exchanged primary and all of their Virasoro descendants into one object.  The contribution of each of these, then, captures different aspects of the dynamics of the theory. For instance, the vacuum Virasoro conformal blocks repackage all multi-stress tensor exchanges which, in the holographic context, relates them to Renyi entropies~\cite{Hartman:2013mia}, thermalization in black hole backgrounds~\cite{Fitzpatrick:2015zha}, or generally with all of the graviton exchanges in AdS. Virasoro conformal blocks are also relevant for the construction of the Hilbert space of $3d$ gravity~\cite{Verlinde:1989ua,Collier:2023fwi,Collier:2025lux}.

Despite their importance, general analytical expressions for the Virasoro conformal blocks have historically been elusive, even for the simplest case of four-point correlators on the Riemann sphere.  To date, the most effective approaches to work with these conformal blocks have been to use either the recursion relations pioneered by Zamolodchikov~\cite{Zamolodchikov:1984eqp,Zamolodchikov:1987avt,Perlmutter:2015iya},\footnote{See~\cite{Cho:2017oxl} for the extension of this technique to the case of higher-point correlation functions.} or the combinatorial expressions arising from the AGT correspondence~\cite{Alday:2009aq,Alba:2010qc}.  

The aim of the present paper is to provide new analytic expressions for both the Virasoro OPE as well as four-point Virasoro conformal blocks on the sphere.  This is achieved using the resolution of the identity recently determined in~\cite{Fortin:2024xir}.  The expression of the OPE~\eqref{eq:OPE} is organized nicely in terms of singular vector operators $L_{\expval{r,s}}$ defined in~\eqref{eq:SVcoefficients}, and highlights how the existence of degenerate Verma modules and singular vectors affects directly also the structure of nondegenerate modules. The coefficients appearing in the OPE are determined explicitly by requiring compatibility with the Virasoro fusion rules.  These same coefficients appear in the expression for the Virasoro conformal blocks~\eqref{eq:Virasoro_blocks}, which is completely explicit and only depends on the conformal block labels, as well as the well-known conformal weights of the singular vectors of the Virasoro algebra. The structure of the conformal block expression is reminiscent of that obtained using Zamolodchikov's $h$-recursion relation in that it exhibits the same type of combinatorial sums, but is organized in terms of the level of descendants.  The conformal block expression can also be evaluated explicitly in certain global and semi-classical limits, reproducing known results.

This paper is organized as follows: Section~\ref{sec:resolution_identity} reviews the Virasoro algebra as well as the resolution of the identity.  It then introduces the Virasoro OPE in terms of generalized Pochhammer symbols, that are simplified with the help of the fusion rules from their original form as a sum over partitions to a trivial product.  Section~\ref{sec:Virasoro_blocks} provides a new explicit expression for the Virasoro conformal blocks from the sewing procedure that is subsequently compared with Zamolodchikov's $h$-recursion relation.  We argue that our result is a genuinely new expression and that it is as simple/complicated as Zamolodchikov's when considering the complexity of the summations.  Several limits (global, semi-classical) and special cases (mixed of heavy and light fields) are then discussed in Section~\ref{sec:limits_blocks}.  Finally, four appendixes gather several results on the Virasoro OPE and its limits (Appendix~\ref{sec:app_OPE}), the Virasoro conformal blocks and double Gamma functions (Appendix~\ref{sec:app_DoubleGammaFunction}) ubiquitous in Liouville theory, a proof relevant to the results of Section~\ref{sec:limits_blocks} (Appendix~\ref{app:higher-r_subleading}), as well as a derivation of the Virasoro Casimir differential operators for four-point Virasoro blocks on the sphere showing that the latter are eigenfunctions of the former with appropriate eigenvalues (Appendix~\ref{app:Casimir}).


\section{Resolution of the identity and Virasoro OPE}\label{sec:resolution_identity}

This section describes our notation and quickly reviews the resolution of the identity obtained from the inverse Shapovalov form derived in~\cite{Fortin:2024xir}.  We also introduce the Virasoro OPE and re-express a ubiquitous quantity, first defined in terms of a sum over standard partitions of the singular vector coefficients times generalizations of the Pochhammer symbol, as a simple product constrained by the fusion rules~\cite{Belavin:1984vu}.


\subsection{Virasoro algebra}

The Virasoro algebra, with generators $L_n$ for $n\in \mathbb{Z}$ and central element $\hat{c}$, is an infinite-dimensional Lie algebra defined by
\begin{equation}
\begin{gathered}
\comm{L_m}{L_n}=(m-n)L_{m+n}+\frac{\hat{c}}{12}(m^3-m)\delta_{m+n,0}\,,\\
\comm{\hat{c}}{L_n}=0\,.
\label{eq:Virasoro_algebra}
\end{gathered}
\end{equation}
The Verma modules of the Virasoro algebra, denoted by $V(c,h)$, are generated by the action of the raising operators $L_{n<0}$ on the lowest-weight vector $\ket{c,h}$ satisfying
\begin{equation}
\hat{c}\ket{c,h}=c\ket{c,h}\,,\qquad L_0\ket{c,h}=h\ket{c,h}\,,\qquad L_{n>0}\ket{c,h}=0\,,
\end{equation}
where $c$ is the central charge of the CFT and $h$ is the conformal dimension of the primary field $\phi(z)$ corresponding to the lowest-weight state $\phi(0)\ket{0}\equiv\ket{c,h}$ through the state-operator correspondence.  A basis for (generic) Verma modules is provided by the descendants
\begin{equation}
L_{-\mu}\ket{h}\equiv L_{-\mu_{l_\mu}}\cdots L_{-\mu_1}\ket{h}\,,
\end{equation}
with $\mu$ ranging over all standard partitions of positive integers and where, for compactness, we introduced $\ket{h}\equiv\ket{c,h}$.  A standard partition is a set of strictly-positive integers $\mu=(\mu_1,\cdots,\mu_{l_\mu})$ with $1\leq\mu_1\leq\cdots\leq\mu_{l_\mu}$, length $l_\mu$, and level $|\mu|=\sum_{i=1}^{l_\mu}\mu_i$.  In the following, we will find it convenient to express the central charge in terms of the parameter $b$,
\begin{equation}
c=1+6\left(b+\frac{1}{b}\right)^2\,,
\label{eq:centralcharge}
\end{equation}
and the conformal dimensions $h_i$ as functions of the momenta $P_i$,
\begin{equation}
h_i(P_i)=\frac{1}{4}\left(b+\frac{1}{b}\right)^2-P_i^2\,.
\label{eq:hP}
\end{equation}

From the Shapovalov form $[S_\ell(c,h)]_{\mu\nu}=\bra{h}L_\mu L_{-\nu}\ket{h}$, constructed as a Gram matrix for a Verma module with $\bra{h}\equiv(\ket{h})^\dagger$, $L_\mu\equiv(L_{-\mu})^\dagger$, and $\expval{h|h}=1$, the Kac determinant~\cite{Kac:1978ge} with $p(\ell)$ the number of partitions at level $\ell$
\begin{equation}
\det S_\ell(c,h)\propto\prod_{\substack{r,s\geq1\\1\leq rs\leq\ell}}(h-h_{\expval{r,s}})^{p(\ell-rs)}\,,
\label{eq:Kac}
\end{equation}
exemplifies the fact that Verma modules are irreducible representations of the Virasoro algebra unless their conformal dimension (or their momentum) are
\begin{equation}
h_{\expval{r,s}}= \frac{b^2(r+1)+(s+1)}{2}\frac{b^{-2}(1-s)+(1-r)}{2}\,,\qquad P_{\expval{r,s}}=\frac{b\,r+b^{-1}s}{2}\,,
\label{eq:hPSV}
\end{equation}
for any strictly-positive integers $r$ and $s$.  In such a case, the associated Verma module $V(c,h_{\expval{r,s}})$ has a singular vector at level $rs$ given by $L_{\expval{r,s}}\!\ket{h_{\expval{r,s}}}$ where
\begin{equation}
L_{\expval{r,s}}=v_{\expval{r,s}}^\mu L_{-\mu}\,.
\label{eq:SVcoefficients}
\end{equation}
The singular vector operators $L_{\expval{r,s}}$ appearing in~\eqref{eq:SVcoefficients}, which in an abuse of notation we will also refer to as singular vectors, are normalized conventionally as $v_{\expval{r,s}}^{(1,\cdots,1)}=1$.


\subsection{Resolution of the identity}

In~\cite{Fortin:2024xir}, we proved an explicit expression for the inverse Shapovalov form,
\begin{equation}
\mathbf{S}_\ell^{-1}(\hat{c},L_0)\equiv L_{-\mu}\left[S_\ell^{-1}(\hat{c},L_0)\right]^{\mu\nu}L_\nu\,,
\label{eq:inv_Shapovalov}
\end{equation}
in terms of the singular vectors~\eqref{eq:SVcoefficients} and their conformal dimensions~\eqref{eq:hPSV}, leading to the following resolution of the identity for the Verma module $V(c,h)$:
\begin{equation}
\mathds{1}(c,h)=\sum_{\ell\geq0}L_{-\mu}\ket{h}\!\bra{h}\left[S_\ell^{-1}(\hat{c},L_0)\right]^{\mu\nu}L_\nu=\sum_{\substack{\ell\geq0\\\boldsymbol{r}\cdot\boldsymbol{s}=\ell}}L_{\expval{\boldsymbol{r},\boldsymbol{s}}}\ket{h}\!\bra{h}\frac{q_{\expval{\boldsymbol{r},\boldsymbol{s}}}}{L_0-h_{\expval{r_1,s_1}}}L_{\expval{\boldsymbol{r},\boldsymbol{s}}}^\dagger\,.
\label{eq:resolution_identity}
\end{equation}
In the final equality, the sum at fixed level $\ell>0$ is over all pairs of $m$-vectors of strictly positive integers $\boldsymbol{r}=(r_1,\cdots,r_m)$ and $\boldsymbol{s}=(s_1,\cdots,s_m)$ with $m$ arbitrary and such that their scalar product matches the level. The $\ell=0$ contribution is understood to be $\ket{h}\!\bra{h}$.  The remaining quantities are defined by
\begin{equation}
L_{\expval{\boldsymbol{r},\boldsymbol{s}}}\equiv L_{\expval{r_m,s_m}}\cdots L_{\expval{r_1,s_1}}\,,\qquad L_{\expval{\boldsymbol{r},\boldsymbol{s}}}^\dagger\equiv L_{\expval{r_1,s_1}}^\dagger\cdots L_{\expval{r_m,s_m}}^\dagger\,,
\label{eq:Lrs}
\end{equation}
for products of singular vectors, as well as
\begin{equation}
q_{\expval{\boldsymbol{r},\boldsymbol{s}}}\equiv q_{\expval{r_1,s_1}\cdots\expval{r_m,s_m}}=\frac{\prod_{i=1}^mq_{\expval{r_i,s_i}}}{\prod_{j=2}^m\left(h_{\expval{r_{j-1},s_{j-1}}}+r_{j-1}s_{j-1}-h_{\expval{r_j,s_j}}\right)}\,,
\label{eq:q_coefficients}
\end{equation}
for the $q$-coefficients, which are functions of the inverse of the regularized norm squared of singular vectors~\cite{Zamolodchikov:2003yb,Yanagida:2010qm}
\begin{equation}
\frac{1}{q_{\expval{r,s}}}\equiv\lim_{h\to h_{\expval{r,s}}}\frac{v_{\expval{r,s}}^\mu[S_{rs}(c,h)]_{\mu\nu}v_{\expval{r,s}}^\nu}{h-h_{\expval{r,s}}}=\frac{2(-1)^{rs+1}}{rs}\prod_{j=1}^r(jb^2)_s(-jb^2)_s\prod_{k=1}^s(k/b^2)_r(-k/b^2)_r\,.
\label{eq:qrs}
\end{equation}
%


\subsection{Determining the Virasoro OPE}\label{ssec:determine_OPE}

The resolution of the identity~\eqref{eq:resolution_identity} with an orthonormal Zamolodchikov metric $\expval{h_i|h_j}=\delta_{ij}$ can be translated directly into the Virasoro OPE for primary fields
\begin{equation}
\tcboxmath{\phi_i(z)\phi_j(0)=\sum_k\bra{h_k}\phi_i(1)\ket{h_j}\sum_{\substack{\ell\geq0\\\boldsymbol{r}\cdot\boldsymbol{s}=\ell}}\frac{q_{\expval{\boldsymbol{r},\boldsymbol{s}}}[h_k-h_j,h_i]_{\expval{\boldsymbol{r},\boldsymbol{s}}}}{h_k-h_{\expval{r_1,s_1}}}\frac{1}{z^{h_i+h_j-h_k-\ell}}L_{\expval{\boldsymbol{r},\boldsymbol{s}}}\phi_k(0)\,,}
\label{eq:OPE}
\end{equation}
as is discussed in detail in Appendix~\ref{sec:app_OPE}.
Here we expressed the action of singular vector operators $L_{\expval{\boldsymbol{r},\boldsymbol{s}}}^\dagger$ on a three-point function via the quantities
\begin{equation}
\begin{gathered}
{}[\alpha,\beta]_{\expval{\boldsymbol{r},\boldsymbol{s}}}\equiv\prod_{j=1}^m\left[\alpha+\sum_{k=1}^{j-1}r_ks_k,\beta\right]_{\expval{r_j,s_j}}\,,\\
[\alpha,\beta]_{\expval{r,s}}\equiv v_{\expval{r,s}}^\mu[\alpha,\beta]_\mu\,,
\end{gathered}
\label{eq:abrs}
\end{equation}
which are written in terms of a generalized Pochhammer symbol,
\begin{equation}
[\alpha,\beta]_\mu\equiv\prod_{j=1}^{l_\mu}\left[\alpha+\sum_{k=1}^{j-1}\mu_k+\beta\mu_j\right]\,,
\label{eq:gen_Pochhammer}
\end{equation}
defined for two numbers $\alpha$ and $\beta$ and one partition $\mu=(\mu_1,\cdots,\mu_{l_\mu})$, while the $\ell=0$ contribution to the last sum in~\eqref{eq:OPE} is simply $1/z^{h_i+h_j-h_k}$. From now on $[\alpha,\beta]_{\mu}$ will appear infrequently so we will refer to $[\alpha,\beta]_{\expval{\boldsymbol r,\boldsymbol s}}$ as generalized Pochhammers with no risk of confusion.

Both the resolution of the identity~\eqref{eq:resolution_identity} and the Virasoro OPE~\eqref{eq:OPE} are momentarily defined with respect to the singular vector coefficients $v_{\expval{r,s}}^\mu$ through the singular vectors~\eqref{eq:SVcoefficients} and through the generalized Pochhammer symbols~\eqref{eq:abrs}.  Even though exact expressions for $v_{\expval{r,s}}^\mu$ exist~\cite{Benoit:1988aw,Bauer:1991ai,Millionshchikov2016,Watts:2024pnm}, they are given in an overcomplete basis and, as such, their use would be somewhat impractical due to proliferation of terms.  In fact, even if the singular vector coefficients were known in the standard partition basis, the number of terms in the sum $[\alpha,\beta]_{\expval{r,s}}$ would increase exponentially with the square root of the level $rs$.  As a consequence, computations of conformal blocks, either from the sewing procedure~\cite{Alessandrini:1971dd,Sonoda:1988mf} with the resolution of the identity or directly from the OPE, would be inefficient.

Fortunately, we can bypass this problem by leveraging the fusion rules~\cite{Belavin:1984vu}.  Indeed, from the Virasoro OPE~\eqref{eq:OPE}, it is clear that some contributions to the three-point correlation function $\expval{\phi_{\expval{r,s}}(w)\phi_i(z)\phi_j(0)}$ diverge unless the associated OPE coefficient $\bra{h_{\expval{r,s}}}\phi_i(1)\ket{h_j}$ vanishes or the fusion rules are satisfied.  The latter case corresponds to when the conformal dimensions, expressed in terms of their momenta~\eqref{eq:hP} and~\eqref{eq:hPSV}, satisfy $P_i\pm P_j=P_{\expval{-r+1+2m,-s+1+2n}}$ for $m\in\{0,\cdots,r-1\}$ and $n\in\{0,\cdots,s-1\}$~\cite{Belavin:1984vu}.  Consequently, to eliminate the divergences originating from the poles in~\eqref{eq:OPE} for primary fields satisfying the fusion rules, we must have
\begin{equation}
[h_{\expval{r,s}}-h_j,h_i]_{\expval{r,s}}=\hspace*{-4pt}\prod_{\substack{0\leq m\leq r-1\\0\leq n\leq s-1}}\hspace*{-4pt}\left(P_j+P_i-P_{\expval{-r+1+2m,-s+1+2n}}\right)\left(P_j-P_i+P_{\expval{-r+1+2m,-s+1+2n}}\right)\,,
\label{eq:hhhrs}
\end{equation}
where the product~\eqref{eq:hhhrs} exhausts the $h_i$ and $h_j$ dependence of the generalized Pochhammer symbol written as $v_{\expval{r,s}}^\mu[h_{\expval{r,s}}-h_j,h_i]_\mu$.  The overall proportionality constant can be set by comparing the coefficients of the highest powers of $h_i$ and $h_j$ in the product~\eqref{eq:hhhrs}, which are contained in $(h_i-h_j)^{rs}$, with the same coefficients computed from the generalized Pochhammer symbol~\eqref{eq:abrs}
\begin{equation}
\begin{aligned}
{}[h_{\expval{r,s}}-h_j,h_i]_{\expval{r,s}}&=v_{\expval{r,s}}^\mu[h_{\expval{r,s}}-h_j,h_i]_\mu=v_{\expval{r,s}}^{(1,\cdots,1)}[h_{\expval{r,s}}-h_j,h_i]_{(1,\cdots,1)}+\cdots\\
&=(h_{\expval{r,s}}-h_j+h_i)_{rs}+\cdots=(h_i-h_j)^{rs}+\cdots\,.
\end{aligned}
\end{equation}
The comparison can be done remembering that $v_{\expval{r,s}}^{(1,\cdots,1)}=1$ and noticing that the generalized Pochhammer symbol for the $(1,\cdots,1)$ partition at level $\ell$, which is exactly the standard Pochhammer symbol $[\alpha,\beta]_{(1,\cdots,1)}=(\alpha+\beta)_\ell$, generates the highest power of $\alpha$ and $\beta$ at that level.

With~\eqref{eq:hhhrs} in hand, we can straightforwardly replace the inefficient sum over standard partitions in $[\alpha,\beta]_{\expval{r,s}}=v_{\expval{r,s}}^\mu[\alpha,\beta]_\mu$~\eqref{eq:abrs} by the simple product
\begin{equation}
\begin{aligned}
{}[\alpha,\beta]_{\expval{r,s}}&=\prod_{\substack{0\leq m\leq r-1\\0\leq n\leq s-1}}\left[\alpha+P_{\expval{r,s}}^2-\left(\sqrt{P_{\expval{1,1}}^2-\beta}-P_{\expval{-r+1+2m,-s+1+2n}}\right)^2\right]\\
&=\!\prod_{\substack{0\leq m\leq r-1\\0\leq n\leq s-1}}\!\left[\alpha-\!\left(\sqrt{P_{\expval{1,1}}^2-\beta}-P_{\expval{2m+1,2n+1}}\right)\!\left(\sqrt{P_{\expval{1,1}}^2-\beta}+P_{\expval{2r-1-2m,2s-1-2n}}\right)\right],
\end{aligned}
\label{eq:Pochhammerrs}
\end{equation}
where the second equality is obtained by carefully recombining the different factors in the first equality. We will mainly be interested in the cases where $\alpha=h_k-h_j+\ell$ and the conformal dimensions are expressed in terms of momenta~\eqref{eq:hP}, leading to the very simple formula:
\begin{equation}
[h_k-h_j+\ell,h_i]_{\expval{r,s}}=\prod_{\substack{0\leq m\leq r-1\\0\leq n\leq s-1}}\left[P_j^2-P_k^2+\ell-\left(P_i-P_{\expval{2m+1,2n+1}}\right)\left(P_i+P_{\expval{2r-1-2m,2s-1-2n}}\right)\right].
\label{eq:gen_Poch_explicit}
\end{equation}
Note that, despite the notation with momenta, these are still polynomial expressions in the conformal weights, as one can check by combining the $(m,n)$ term with the $(r-1-m,s-1-n)$ term.  The determination of these coefficients makes the expression of the Virasoro OPE~\eqref{eq:OPE} explicit, and will allow us to write explicit expressions for the four-point Virasoro conformal blocks on the sphere, to which we now turn.


\section{Four-point Virasoro blocks on the sphere}\label{sec:Virasoro_blocks}

In this section we compute Virasoro blocks on the sphere from the resolution of the identity (or, equivalently, the Virasoro OPE) and compare our result to the celebrated Zamolodchikov $h$-recursion relation.


\subsection{Four-point Virasoro blocks on the sphere}

The sewing procedure~\cite{Alessandrini:1971dd,Sonoda:1988mf} re-expresses arbitrary correlation functions in $2d$ CFTs in terms of sums of smaller correlation functions through the introduction of the resolution of the identity. Focusing on one chiral half of the correlator, the sewing procedure applied to four-point correlators on the sphere dictates that
\begin{equation}
\expval{\phi_1(z_1)\phi_2(z_2)\phi_3(z_3)\phi_4(z_4)}=\sum_h\bra{0}\phi_1(z_1)\phi_2(z_2)\mathds{1}(c,h)\phi_{3}(z_{3})\phi_4(z_4)\ket{0}\,,
\label{eq:sewing}
\end{equation}
where $\mathds{1}(c,h)$ projects on the Verma module $V(c,h)$ and resolves the identity as in~\eqref{eq:resolution_identity}. 
Let us extract a conventional conformal-covariant prefactor
\begin{equation}
\expval{\phi_1(z_1)\phi_2(z_2)\phi_3(z_3)\phi_4(z_4)}=\frac{1}{z_{12}^{h_1+h_2}z_{34}^{h_3+h_4}}\left(\frac{z_{24}}{z_{14}}\right)^{h_{12}}\left(\frac{z_{14}}{z_{13}}\right)^{h_{34}}G(\cratio)\,,
\label{eq:4ptG}
\end{equation}
where we introduced the notation $z_{ij}\equiv z_i-z_j$, $h_{ij}\equiv h_i-h_j$, and where $\cratio\equiv\frac{z_{12}z_{34}}{z_{13}z_{24}}$ is the four-point cross-ratio. The sewing procedure~\eqref{eq:sewing} (or, equivalently, the Virasoro OPE) applied to~\eqref{eq:4ptG} enforces the conformal block decomposition
\begin{equation}
G(\cratio)=\sum_h\bra{h_1}\phi_2(1)\ket{h}\!\bra{h}\phi_3(1)\ket{h_4}\mathcal{F}_h^{\boldsymbol{h}}(\cratio)\,,
\label{eq:block_decomposition}
\end{equation}
where $\boldsymbol{h}=(h_1,h_2,h_3,h_4)$ and the $\mathcal{F}_h^{\boldsymbol{h}}(\cratio)$ are the four-point Virasoro blocks on the sphere, defined as 
\begin{equation}
\mathcal{F}_h^{\boldsymbol{h}}(\cratio)=\lim_{\substack{z_2\to1\\z_3\to\cratio}}z_3^{h_3+h_4}\frac{\bra{h_1}\phi_2(z_2)\mathds{1}(c,h)\phi_{3}(z_{3})\ket{h_4}}{\bra{h_1}\phi_2(1)\ket{h}\!\bra{h}\phi_3(1)\ket{h_4}}\,.
\label{eq:Virasoro_blocks_id}
\end{equation}
At this point, we can insert the resolution of the identity~\eqref{eq:resolution_identity} into~\eqref{eq:Virasoro_blocks_id} to get
\begin{equation}
\mathcal{F}_h^{\boldsymbol{h}}(\cratio)=\cratio^{h_3+h_4}\lim_{\substack{z_2\to1\\z_3\to\cratio}}\sum_{\substack{\ell\geq0\\\boldsymbol{r}\cdot\boldsymbol{s}=\ell}}\frac{q_{\expval{\boldsymbol{r},\boldsymbol{s}}}}{h-h_{\expval{r_1,s_1}}}\frac{\bra{h_1}\phi_2(z_2)L_{\expval{\boldsymbol{r},\boldsymbol{s}}}\ket{h}}{\bra{h_1}\phi_2(1)\ket{h}}\frac{\bra{h}L_{\expval{\boldsymbol{r},\boldsymbol{s}}}^\dagger\phi_{3}(z_{3})\ket{h_4}}{\bra{h}\phi_3(1)\ket{h_4}}\,,
\end{equation}
where we can finally explicitly evaluate the three-point functions via the identities
\begin{equation*}
\begin{gathered}
\frac{\bra{h_j}\phi_i(z)L_{\expval{\boldsymbol{r},\boldsymbol{s}}}\ket{h_k}}{\bra{h_j}\phi_i(1)\ket{h_k}}=\mathcal{L}_{\expval{\boldsymbol{r},\boldsymbol{s}}}\frac{\bra{h_j}\phi_i(z)\ket{h_k}}{\bra{h_j}\phi_i(1)\ket{h_k}}=\mathcal{L}_{\expval{\boldsymbol{r},\boldsymbol{s}}}\frac{1}{z^{h_i-h_j+h_k}}=\frac{[h_k-h_j,h_i]_{\expval{\boldsymbol{r},\boldsymbol{s}}}}{z^{h_i-h_j+h_k+rs}}\,,\\
\frac{\bra{h_k}L_{\expval{\boldsymbol{r},\boldsymbol{s}}}^\dagger\phi_i(z)\ket{h_j}}{\bra{h_k}\phi_i(1)\ket{h_j}}=\mathcal{L}_{\expval{\boldsymbol{r},\boldsymbol{s}}}^\dagger\frac{\bra{h_k}\phi_i(z)\ket{h_j}}{\bra{h_k}\phi_i(1)\ket{h_j}}=\mathcal{L}_{\expval{\boldsymbol{r},\boldsymbol{s}}}^\dagger\frac{1}{z^{h_i+h_j-h_k}}=\frac{[h_k-h_j,h_i]_{\expval{\boldsymbol{r},\boldsymbol{s}}}}{z^{h_i+h_j-h_k-rs}}\,,
\end{gathered}
\end{equation*}
which we work out in~\eqref{eq:3ptL}.

This leads to the second main result of this paper: the explicit expression for four-point Virasoro conformal blocks on the sphere
\begin{equation}
\tcboxmath{\mathcal{F}_h^{\boldsymbol{h}}(\cratio)=\sum_{\substack{\ell\geq0\\\boldsymbol{r}\cdot\boldsymbol{s}=\ell}}\frac{q_{\expval{\boldsymbol{r},\boldsymbol{s}}}}{h-h_{\expval{r_1,s_1}}}[h-h_1,h_2]_{\expval{\boldsymbol{r},\boldsymbol{s}}}[h-h_4,h_3]_{\expval{\boldsymbol{r},\boldsymbol{s}}}\cratio^{h+\ell}\,,}
\label{eq:Virasoro_blocks}
\end{equation}
where the $\ell=0$ contribution is given by $\cratio^h$. We remind the reader that for every $\ell>0$ the sum runs over all sets $\expval{\boldsymbol{r},\boldsymbol{s}}=\{\expval{r_1,s_1},\dots,\expval{r_m,s_m}\}$ of total level $\ell=\sum_{k=1}^mr_ks_k$, that the $q_{\expval{\boldsymbol{r},\boldsymbol{s}}}$ are defined by~\eqref{eq:q_coefficients}--\eqref{eq:qrs}, and that 
\begin{equation}
[h-h_j,h_i]_{\expval{\boldsymbol{r},\boldsymbol{s}}}=\prod_{a=1}^m\left[h-h_j+\sum_{k=1}^{a-1}r_ks_k,h_i\right]_{\expval{r_a,s_a}}\,,
\end{equation}
with factors determined in~\eqref{eq:gen_Poch_explicit}. 

Expression~\eqref{eq:Virasoro_blocks} is currently among the most efficient ways to compute the series expansion in the standard cross-ratio $\cratio$ of generic four-point Virasoro blocks. One of its main strengths is, in fact, the high degree of reuse of lower-level coefficients in the expression of the higher-level ones. In an ancillary file of this work, we include a \textit{Mathematica} notebook with a simple implementation of the main formula~\eqref{eq:Virasoro_blocks}.  We will discuss in the next subsection how our expression has the same sum complexity as Zamolodchikov's $h$-recursion formula, while not depending explicitly on the behavior at $h\to\infty$.


\subsection{Comparison with Zamolodchikov's \texorpdfstring{$h$}{h}-recursion relation}\label{ssec:Zamolodchikov_comparison}

Our next step is to compare our expression for the conformal blocks~\eqref{eq:Virasoro_blocks} with the celebrated $h$-recursion relation of Zamolodchikov~\cite{Zamolodchikov:1984eqp,Zamolodchikov:1987avt}.  As we will see, this will provide another proof of the identity~\eqref{eq:gen_Poch_explicit} for the generalized Pochhammer symbols $[\alpha,\beta]_{\expval{r,s}}$ that are used in our definition of the blocks.

Given the expression of the conformal blocks~\eqref{eq:Virasoro_blocks_id} in terms of the inverse Shapovalov form through the resolution of the identity~\eqref{eq:resolution_identity}, together with the knowledge of all zeros of the Kac determinant~\eqref{eq:Kac}, one can write a Mittag-Leffler expansion in either the poles in the central charge $c$ or the conformal dimension $h$.  This is the starting point of Zamolodchikov's famous recursion relations\footnote{Technically, Zamolodchikov's $h$-recursion relation is not a Mittag-Leffler expansion of the conformal blocks $\mathcal{F}_h^{\boldsymbol{h}}(\cratio)$ as in~\cite{Zamolodchikov:1984eqp}.  It is a Mittag-Leffler expansion of the ratio $\mathcal{F}_h^{\boldsymbol{h}}(\cratio)/f_h^{\boldsymbol{h}}(\cratio)$ where $f_h^{\boldsymbol{h}}(x)$ is defined in~\eqref{eq:f}~\cite{Zamolodchikov:1987avt}.} that can be used to efficiently compute the Virasoro blocks~\cite{Zamolodchikov:1984eqp,Zamolodchikov:1987avt}.  Given the explicit appearance of the poles in $h$ in our expression for the conformal blocks~\eqref{eq:Virasoro_blocks}, it is natural to compare this with the $h$-recursion of Zamolodchikov~\cite{Zamolodchikov:1987avt}, which reads
\begin{equation}
\mathcal{F}_h^{\boldsymbol{h}}(\cratio)=f_h^{\boldsymbol{h}}(\cratio)+\sum_{r,s>0}\frac{\mathcal{R}_{\expval{r,s}}}{h-h_{\expval{r,s}}}(16\mathfrak{q})^{h-h_{\expval{r,s}}}\mathcal{F}_{h_{\expval{r,s}}+rs}^{\boldsymbol{h}}(\cratio)
\,.
\label{eq:Zamolodchikov_recursion}
\end{equation}
Here, the $h$-residues are proportional to conformal blocks of the corresponding submodules in the degenerate Verma module $V(c,h_{\expval{r,s}})$ and the function $f_h^{\boldsymbol{h}}(\cratio)$ encodes the large-$h$ behavior of the conformal blocks~\cite{Zamolodchikov:1984eqp,Zamolodchikov:1987avt}.

More specifically, on the one hand the coefficients $\mathcal{R}_{\expval{r,s}}$ appearing in the residues of~\eqref{eq:Zamolodchikov_recursion} are given by
\begin{equation}
\mathcal{R}_{\expval{r,s}}=q_{\expval{r,s}}\left[h_{\expval{r,s}}-h_1,h_2\right]_{\expval{r,s}}\left[h_{\expval{r,s}}-h_4,h_3\right]_{\expval{r,s}}\,,
\end{equation}
in our notation, using~\eqref{eq:gen_Poch_explicit}.  On the other hand, $f_h^{\boldsymbol{h}}(\cratio)$ is a function of the cross-ratio expressed as
\begin{equation}
f_h^{\boldsymbol{h}}(\cratio)=\cratio^h(1-\cratio)^{\frac{c-1}{24}-h_2-h_3}\left(\frac{16\mathfrak{q}}{\cratio}\right)^{h-\frac{c-1}{24}}[\theta_3(\mathfrak{q})]^{\frac{c-1}{2}-4(h_1+h_2+h_3+h_4)}\,,\label{eq:f}
\end{equation}
where $\theta_3(\mathfrak{q})$ and $\mathfrak{q}$, given by
\begin{equation}
\begin{gathered}
\theta_3(\mathfrak{q})=\sum_{n\in\mathbb{Z}}\mathfrak{q}^{n^2}=1+2\mathfrak{q}+2\mathfrak{q}^4+\cdots\,,\\
\mathfrak{q}=\exp\left[-\pi\frac{{}_2F_1(1/2,1/2;1;1-\cratio)}{{}_2F_1(1/2,1/2;1;\cratio)}\right]=\frac{\cratio}{16}+\frac{\cratio^2}{32}+\cdots\,,
\end{gathered}
\label{eq:nome}
\end{equation}
are a theta function and the so-called nome, respectively~\cite{Zamolodchikov:1987avt}.

To compare with this, we can straightforwardly compute the residues of our expression for the Virasoro blocks~\eqref{eq:Virasoro_blocks} as
\begin{equation}
\begin{aligned}
\underset{\,\,\,h=h_{\expval{r,s}}}{\mathrm{Res}}\hspace*{-7pt}\mathcal{F}_h^{\boldsymbol{h}}(\cratio)&=q_{\expval{r,s}}\left[h_{\expval{r,s}}-h_1,h_2\right]_{\expval{r,s}}\left[h_{\expval{r,s}}-h_4,h_3\right]_{\expval{r,s}}\sum_{\substack{\ell\geq0\\\boldsymbol{r}\cdot\boldsymbol{s}=\ell}}\frac{q_{\expval{r,s}\expval{\boldsymbol{r},\boldsymbol{s}}}}{q_{\expval{r,s}}}\\
&\qquad\times\left[h_{\expval{r,s}}+rs-h_1,h_2\right]_{\expval{\boldsymbol{r},\boldsymbol{s}}}\left[h_{\expval{r,s}}+rs-h_4,h_3\right]_{\expval{\boldsymbol{r},\boldsymbol{s}}}\cratio^{h_{\expval{r,s}}+rs+\ell}\\
&=q_{\expval{r,s}}\left[h_{\expval{r,s}}-h_1,h_2\right]_{\expval{r,s}}\left[h_{\expval{r,s}}-h_4,h_3\right]_{\expval{r,s}}\mathcal{F}_{h_{\expval{r,s}}+rs}^{\boldsymbol{h}}(\cratio)\,,
\end{aligned}
\end{equation}
where we used the definition~\eqref{eq:abrs} for the generalized Pochhammer symbols and~\eqref{eq:q_coefficients} to check that
\begin{equation}
\frac{q_{\expval{r,s}\expval{\boldsymbol{r},\boldsymbol{s}}}}{q_{\expval{r,s}}}=\frac{q_{\expval{\boldsymbol{r},\boldsymbol{s}}}}{h_{\expval{r,s}}+rs-h_{\expval{r_1,s_1}}}\,.
\end{equation}
Hence our formula~\eqref{eq:Virasoro_blocks} has the same residues for the poles in the conformal dimension as~\eqref{eq:Zamolodchikov_recursion}.  We have thus shown that we can explicitly recover the form of Zamolodchikov's $h$-recursion with the correct residues using our expression for the Virasoro conformal blocks.  We stress, however, that our expression~\eqref{eq:Virasoro_blocks} is very different in nature to Zamolodchikov's recursion.  We can in fact decompose~\eqref{eq:Virasoro_blocks} as
\begin{equation}
\begin{aligned}
\mathcal{F}_h^{\boldsymbol{h}}(\cratio)&=\cratio^h+\sum_{r,s>0}\frac{\mathcal{R}_{\expval{r,s}}}{h-h_{\expval{r,s}}}\left\{\frac{[h-h_1,h_2]_{\expval{r,s}}}{[h_{\expval{r,s}}-h_1,h_2]_{\expval{r,s}}}\frac{[h-h_4,h_3]_{\expval{r,s}}}{[h_{\expval{r,s}}-h_4,h_3]_{\expval{r,s}}}\right.\\
&\qquad\times\left.\sum_{\substack{\ell\geq0\\\boldsymbol{r}\cdot\boldsymbol{s}=\ell}}\frac{q_{\expval{\boldsymbol{r},\boldsymbol{s}}}}{h_{\expval{r,s}}+rs-h_{\expval{r_1,s_1}}}[h+rs-h_1,h_2]_{\expval{\boldsymbol{r},\boldsymbol{s}}}[h+rs-h_4,h_3]_{\expval{\boldsymbol{r},\boldsymbol{s}}}\cratio^{h+rs+\ell}\right\}\,,
\end{aligned}
\end{equation}
where it is clear that the object in curly brackets correspond to $\mathcal{F}_{h_{\expval{r,s}}+rs}^{\boldsymbol{h}}(\cratio)$ only when $h$ is set to $h_{\expval{r,s}}$, matching the residues of~\eqref{eq:Zamolodchikov_recursion} as already proven.  Hence~\eqref{eq:Virasoro_blocks} is neither Zamolodchikov's $h$-recursion relation nor can it be put into a simple recursion relation as in~\cite{Zamolodchikov:1987avt}.

Despite their different nature, both expressions share the same complexity in sums.  Indeed, the solution of~\eqref{eq:Zamolodchikov_recursion} can be written in our notation as
\begin{equation}
\begin{aligned}
\mathcal{F}_h^{\boldsymbol{h}}(\cratio)&=\sum_{\substack{\ell\geq0\\\boldsymbol{r}\cdot\boldsymbol{s}=\ell}}\frac{q_{\expval{\boldsymbol{r},\boldsymbol{s}}}}{h-h_{\expval{r_1,s_1}}}\left[\prod_{i\geq1}^m[h_{\expval{r_i,s_i}}-h_1,h_2]_{\expval{r_i,s_i}}[h_{\expval{r_i,s_i}}-h_4,h_3]_{\expval{r_i,s_i}}\right]f_{h+\ell}^{\boldsymbol{h}}(\cratio)\\
&=f_h^{\boldsymbol{h}}(\cratio)\sum_{\substack{\ell\geq0\\\boldsymbol{r}\cdot\boldsymbol{s}=\ell}}\frac{q_{\expval{\boldsymbol{r},\boldsymbol{s}}}}{h-h_{\expval{r_1,s_1}}}\left[\prod_{i\geq1}^m[h_{\expval{r_i,s_i}}-h_1,h_2]_{\expval{r_i,s_i}}[h_{\expval{r_i,s_i}}-h_4,h_3]_{\expval{r_i,s_i}}\right](16\mathfrak{q})^\ell\,,
\end{aligned}
\label{eq:Zamolodchikov_soln}
\end{equation}
where the $\ell=0$ contribution is simply $f_h^{\boldsymbol{h}}(\cratio)$.  It is now straightforward to verify from~\eqref{eq:Zamolodchikov_soln} that our result~\eqref{eq:Virasoro_blocks} has the same sum ranges as the solution of Zamolodchikov's $h$-recursion relation.  Note that, despite the use of the same label $\ell$, the ``level'' in~\eqref{eq:Zamolodchikov_soln} and thus in~\eqref{eq:Zamolodchikov_recursion} does not have the same clear meaning as the level in~\eqref{eq:Virasoro_blocks}.

Moreover, our formula~\eqref{eq:Virasoro_blocks} can be directly compared with the previously known general expressions for conformal blocks obtained in~\cite{Perlmutter:2015iya} via both recursion relations of Zamolodchikov.  For the $h$-recursion~\cite[Eq.2.47]{Perlmutter:2015iya}, which should match~\eqref{eq:Zamolodchikov_soln}, one has the same type of structure but with different coefficients and a different expansion variable.  The $c$-recursion~\cite[Eq.2.32]{Perlmutter:2015iya}, instead, has the same expansion variable $\cratio$ as our expression. However, \cite[Eq.2.32]{Perlmutter:2015iya} differs both in its general structure with a less regular sum range, and in the coefficients which involve shifts in the central charge in addition to the shifts in the conformal dimension.

Since other known expressions for the conformal blocks are either: 1) obtained from Zamolodchikov's recursion relations~\cite{Perlmutter:2015iya}; 2) computed from the AGT correspondence with an explicit prefactor in the cross-ratio times a double sum over partitions~\cite{Alday:2009aq,Alba:2010qc}; or 3) derived in special cases such as with external degenerate operators~\cite{Belavin:1984vu,Dotsenko:1984nm,Dotsenko:1984ad,Bissi:2024wur} or in the semi-classical limit~\cite{Zamolodchikov:1995aa,Litvinov:2013sxa,Fitzpatrick:2014vua,Fitzpatrick:2015zha,Perlmutter:2015iya,Beccaria:2015shq,Chen:2016cms,Maloney:2016kee,Kusuki:2018nms,Alkalaev:2019zhs,Alkalaev:2024knk,Bissi:2024wur}; our result~\eqref{eq:Virasoro_blocks} is thus a genuinely new solution for four-point Virasoro blocks on the sphere.

We now move to the analysis of some special limits of conformal blocks derived from our formula.


\section{Limits and special cases}\label{sec:limits_blocks}

In this section, we discuss some simple cases and limits in which our formula~\eqref{eq:Virasoro_blocks} resums to closed-form or asymptotic expressions.  We will start by discussing the simple case of having an identity operator among one of the four external fields, which despite being trivial, allows us to determine practical identities that will be used in deriving other expressions.  We then discuss the Virasoro conformal blocks at large central charge, ranging from the global limit to the leading asymptotics in various semiclassical limits with heavy or light operators. 


\subsection{Consistency with an identity field insertion}\label{ssec:block_identity_limit}

The first case we discuss is when one of the external operators corresponds to the degenerate field with $h=0$, the identity field $ \mathds{1}(z)\equiv \phi_{\expval{1,1}}(z)$.  Since degenerate representations such as that of $\phi_{\expval{1,1}}$ correspond to quotients of two Verma modules---$V(c,0)/V(c,1)$ in this case---it is not enough to just plug $h_i=h_{\expval{r,s}}$ in our formula~\eqref{eq:Virasoro_blocks}.  In the general case, in fact, this procedure would produce a conformal block for the degenerate Verma module $V(c,h_{\expval{r,s}})$, a rather exotic case relevant for theories such as timelike Liouville theory~\cite{Ribault:2015sxa}.

In order to correctly reproduce the conformal block for the field that sits in the representation $V(c,h_{\expval{r,s}})/V(c,h_{\expval{r,s}}+rs)$, one has to enforce the fusion rules at the vertices involving~$\phi_{\expval{r,s}}(z)$.  In the case of the identity $\phi_{\expval{1,1}}(z)$ this simply implies that the other two fields involved in the OPE with it must have equal conformal dimension.

Modulo the $\mathds{Z}_2$ symmetry that exchanges $(h_1,h_2)\leftrightarrow (h_4,h_3)$ in~\eqref{eq:Virasoro_blocks}, there are two different cases to consider: when $h_2=0$, or when $h_1=0$.
\begin{itemize}
    \item If $h_2=0$, the procedure is trivial: the fusion rules enforce
\begin{equation}
    h_2=0 \qquad \Rightarrow \qquad h=h_1\,,
\end{equation}
which reduces the expression of the conformal block to
\begin{equation}
    \mathcal{F}_{h_1}^{(h_1,0,h_3,h_4)}(\cratio)=\sum_{\substack{\ell\geq0\\\boldsymbol{r}\cdot\boldsymbol{s}=\ell}}\frac{q_{\expval{\boldsymbol{r},\boldsymbol{s}}}}{h_1-h_{\expval{r_1,s_1}}}[0,0]_{\expval{\boldsymbol{r},\boldsymbol{s}}}[h_1-h_4,h_3]_{\expval{\boldsymbol{r},\boldsymbol{s}}}\cratio^{h_1+\ell}=\cratio^{h_1}\,,
\end{equation}
since the generalized Pochhammer $[0,0]_{\expval{\boldsymbol{r},\boldsymbol{s}}}$ kills all terms at levels $\ell>0$.  Reintroducing the prefactor extracted in~\eqref{eq:4ptG}, this perfectly reproduces the three-point function
\begin{equation}
    \frac{\left(\frac{z_{24}}{z_{14}}\right)^{h_{1}}\left(\frac{z_{14}}{z_{13}}\right)^{h_{34}}}{z_{12}^{h_1}z_{34}^{h_3+h_4}}\cratio^{h_1}=z_{13}^{h_4-h_1-h_3}z_{14}^{h_{3}-h_1-h_4}z_{34}^{h_1-h_3-h_4}=\frac{\expval{\phi_{1}(z_1)\phi_{3}(z_3)\phi_{4}(z_4)}}{\bra{h_1}\phi_3(1)\ket{h_4}}\,.
\end{equation}
\item If $h_1=0$, the fusion rules enforce
\begin{equation}
    h_1=0 \qquad \Rightarrow \qquad h=h_2\,,
\end{equation}
which, given the non-symmetric way in which $h_1$ and $h_2$ appear in~\eqref{eq:Virasoro_blocks}, leads to a different evaluation.  The conformal block, in fact, becomes
\begin{equation}
    \mathcal{F}_{h_2}^{(0,h_2,h_3,h_4)}(\cratio)=\sum_{\substack{\ell\geq0\\\boldsymbol{r}\cdot\boldsymbol{s}=\ell}}\frac{q_{\expval{\boldsymbol{r},\boldsymbol{s}}}}{h_2-h_{\expval{r_1,s_1}}}[h_2,h_2]_{\expval{\boldsymbol{r},\boldsymbol{s}}}[h_2-h_4,h_3]_{\expval{\boldsymbol{r},\boldsymbol{s}}}\cratio^{h_2+\ell}\,,
    \label{eq:block_identity_1_sum}
\end{equation}
which is non-trivial to evaluate.  However, for everything to be consistent, this must reproduce the three-point correlator $\frac{\expval{\phi_2(z_2)\phi_3(z_3)\phi_4(z_4)}}{\bra{h_2}\phi_3(1)\ket{h_4}}$, which implies that the conformal block must be equal to
\begin{equation}
    \mathcal{F}_{h_2}^{(0,h_2,h_3,h_4)}(\cratio)\stackrel{!}{=}\cratio^{h_2} (1-\cratio)^{h_2+h_{34}} = \cratio^{h_2}\sum_{\ell\ge 0}\frac{(h_2+h_{34})_\ell}{\ell!} \cratio^{\ell}\,.
    \label{eq:block_identity_1_simple}
\end{equation}
Equating~\eqref{eq:block_identity_1_sum} with~\eqref{eq:block_identity_1_simple} level by level, we then obtain the non-trivial identity
\begin{equation}
    \sum_{\boldsymbol{r}\cdot\boldsymbol{s}=\ell}\frac{q_{\expval{\boldsymbol{r},\boldsymbol{s}}}}{h_2-h_{\expval{r_1,s_1}}}[h_2,h_2]_{\expval{\boldsymbol{r},\boldsymbol{s}}}[h_2-h_4,h_3]_{\expval{\boldsymbol{r},\boldsymbol{s}}}= \frac{(h_2+h_{34})_\ell}{\ell!}\,,
    \label{eq:block_identity_2_eq}
\end{equation}
which is easy to check level by level. 
\end{itemize}
Equation~\eqref{eq:block_identity_2_eq} is a powerful identity which we also derive in Appendix~\ref{app:three-point_OPE} directly from the Virasoro OPE.  We can also exploit the freedom in $h_2$, $h_3$, and $h_4$ to derive simpler identities from~\eqref{eq:block_identity_2_eq}.  If we consider the limit of large $h_3$ and extract the leading behavior, we get the beautiful identity
\begin{equation}
\sum_{\boldsymbol{r}\cdot\boldsymbol{s}=\ell}\frac{q_{\expval{\boldsymbol{r},\boldsymbol{s}}}[h,h]_{\expval{\boldsymbol{r},\boldsymbol{s}}}}{h-h_{\expval{r_1,s_1}}}=\frac{1}{\ell!}\,,
\label{eq:OPE_ell!identity} 
\end{equation}
which we derive also in Appendix~\ref{app:IdentityLimitOPE} from the identity limit of the Virasoro OPE.  We can also further take the large $h$ limit in~\eqref{eq:OPE_ell!identity}, which at leading order implies the identity
\begin{equation}
    \sum_{\boldsymbol{r}\cdot\boldsymbol{{s}}=\ell}q_{\expval{\boldsymbol{r},\boldsymbol{s}}}=0\,.
    \label{eq:sum_of_qrs=0}
\end{equation}

All of these identities are non-trivial and can be used to work out conformal blocks in special limits, as we will see in the next subsection.


\subsection{Conformal blocks at large central charge}\label{ssec:large_c}

We now turn to the study of various limits of large central charge for the Virasoro conformal blocks in~\eqref{eq:Virasoro_blocks}.  In the limit of large central charge, the $L_n$ generators of the Virasoro algebra with $\abs{n}>1$ decouple, reducing the Virasoro algebra to the $\mathfrak{sl}(2,\mathds{R})$ algebra of global conformal transformations.  One can consider several situations with large central charge depending on the assumptions of how dimensions of the fields scale alongside the central charge.  If the only large parameter in our expression is the central charge, then the Virasoro conformal blocks~\eqref{eq:Virasoro_blocks} must simply reduce to the global $\mathfrak{sl}(2,\mathds{R})$ blocks.  If, instead, we consider some of the conformal dimensions to scale with the central charge, we are effectively describing via Liouville theory or holography some gravitational system in an almost classical background geometry determined by heavy objects.  These are referred to as semi-classical limits, in which the conformal blocks are expected to exponentiate~\cite{Zamolodchikov1986JETP,Zamolodchikov:1995aa,Besken:2019jyw}.

Throughout this subsection, we will implement the large central charge limit as\footnote{The limit $b\to 0$ is equivalent to the $b\to\infty$ limit that we consider, with the only difference being in the exchanged role of $r$ and $s$ when dealing with the singular vector labels $\expval{r,s}$.} $b\to\infty$ using~\eqref{eq:centralcharge}.  At the level of singular vectors, this limit implies
\begin{equation}
		h_{\expval{r,s}}\stackrel{b\to\infty}{=} \frac{r^2-1}{4} b^2- \frac{rs-1}{2}+\mathrm{O}(b^{-2})\,,
		\label{eq:hrs_scaling_b}
\end{equation}
from which it is evident that the singular vectors with $r>1$ should decouple, leaving at finite conformal dimension only the singular vectors with
\begin{equation}
    h_{\expval{1,s}}\stackrel{b\to\infty}{=}\frac{1-s}{2}+\mathrm{O}(b^{-2})\,,
\end{equation}
that correspond to the singular vectors of the $\mathfrak{sl}(2,\mathds{R})$ algebra and are related to its weight-shifting operators.

In the limits that we will consider, we will distinguish between ``light'' operators with $h_{\text{light}}\stackrel{b\to\infty}{=} \mathrm{O}(b^0)$, and ``heavy'' operators, with $h_{\text{heavy}}\stackrel{b\to\infty}{=} \tilde{h}\,b^2+\mathrm{O}(b^0)$.  In terms of the momenta~\eqref{eq:hP}, these correspond to
\begin{equation}
    P_{\text{light}} \stackrel{b\to\infty}{=} \frac{b}{2}+\frac{1-2h}{2b}+\mathrm{O}(b^{-3})\,,
\end{equation}
and
\begin{equation}
    P_{\text{heavy}}\stackrel{b\to\infty}{=}-\frac{\mathrm{i} \,b}{2}\sqrt{4 \tilde{h}-1}+\mathrm{O}(b^{-1})\,,
\end{equation}
where $P_{\text{heavy}}$ is purely imaginary if the field is above the BTZ threshold $\tilde{h}\ge\frac{1}{4}$ or real if below.\footnote{The sign has been chosen to match $P_{\text{light}}$ in the $\tilde{h}\to0$ limit.}

In what follows, we will examine how each term in the sum in~\eqref{eq:Virasoro_blocks} scales and keep only the leading ones at every level. While this procedure correctly reproduces the complete global block when all fields are light, it will only produce the leading behavior of the exponent in the $\cratio \to 0$ limit for the semiclassical blocks. In other words, our procedure leads to $\exp \!\left(\mathcal{G}\right)$ where $\mathcal{G}\stackrel{b\to\infty}{\simeq}\log \mathcal{F}$ for $\cratio \to 0$. This can be thought of as a proof-of-concept calculation which produces part of the answer, but which would have to be extended to all the positive powers of $b$ that appear at every level in order to compute the complete semiclassical block. 

For our purpose, we rewrite the level-$\ell$ coefficient of the blocks~\eqref{eq:Virasoro_blocks} as
\begin{equation}
    \sum_{\boldsymbol{r}\cdot\boldsymbol{s}=\ell}\frac{\prod_{a=1}^m \left(q_{\expval{r_a,s_a}}\left[h-h_1+\sum_{k=1}^{a-1}r_ks_k,h_2\right]_{\expval{r_a,s_a}}\left[h-h_4+\sum_{k=1}^{a-1}r_ks_k,h_3\right]_{\expval{r_a,s_a}}\right)}{\left(h-h_{\expval{r_1,s_1}}\right) \prod_{j=2}^m\left(h_{\expval{r_{j-1},s_{j-1}}}+r_{j-1}s_{j-1}-h_{\expval{r_j,s_j}}\right)}\,,
    \label{eq:block_coefficient_for_scaling}
\end{equation}
where we expanded the boldfaced $\expval{\boldsymbol{r},\boldsymbol{s}}$ notation.  To indicate the scaling of terms, we will use the symbol $\stackrel{b\to\infty}\sim$, by which we mean to keep the largest power of $b$ and ignore for now factors that are independent of $b$. The symbol $\stackrel{b\to\infty}{\simeq}$ will stand instead for an equality up to sub-leading terms in~$b$. Within the coefficients~\eqref{eq:block_coefficient_for_scaling}, the objects that only depend on the central charge are the inverse Zamolodchikov norms, which scale as
\begin{equation}
    q_{\expval{r_a,s_a}}
    \stackrel{b\to\infty}{\sim}b^{-4(r_a-1)s_a}\,,
    \label{eq:qrs_scaling_b}
\end{equation}
and the combining factors
\begin{equation}
    \prod_{j=2}^m\left(h_{\expval{r_{j-1},s_{j-1}}}+r_{j-1}s_{j-1}-h_{\expval{r_j,s_j}}\right)\sim (b^2)^{\#\text{changes}(r_1,\dots,r_m)}\,,
    \label{eq:combining_factor_scaling_b}
\end{equation}
where $\#\text{changes}(r_1,\dots,r_m)$ counts how many times $r_a\ne r_{a+1}$ in the sequence $(r_1,\dots,r_m)$.  The poles in $h$ scale in a way that depends on $r_1$ if $h$ is light
\begin{equation}
    \left(h-h_{\expval{r_1,s_1}}\right)\stackrel{b\to\infty}\sim \begin{cases}
        b^0 \quad &\text{if} \quad  r_1=1\\
        b^2 \quad &\text{if} \quad r_1>1
    \end{cases}\,,
    \label{eq:poles_scaling_b}
\end{equation}
while their scaling is always the same for heavy operators
\begin{equation}
    \left(\tilde{h}\,b^2-h_{\expval{r_1,s_1}}\right)\stackrel{b\to\infty}\sim b^2\,.
\end{equation}
Finally, the generalized Pochhammers have the most complicated dependence on the conformal dimensions.  For all light fields, we have
\begin{equation}
	\begin{gathered}
		\left[h-h_j+\sum_{k=1}^{a-1}r_ks_k,h_i\right]_{\expval{r_a,s_a}} \stackrel{b\to\infty}{=} \prod_{\substack{0\leq m\leq r_a-1\\0\leq n\leq s_a-1}} \left[m(r_a-m)b^2+ \gamma_{mn} +\mathrm{O}\left(b^{-2}\right)\right]
		\stackrel{b\to\infty}{\sim}b^{2(r_a-1)s_a}\,,\\
		\text{where} \quad \gamma_{mn} = h-h_j+(h_i+n) (r_a-2 m)+m s_a+\sum_{k=1}^{a-1}r_ks_k\,.
	\end{gathered}
	\label{eq:scaling_genPoch_light}
\end{equation}
For heavy fields we distinguish between the generic case
\begin{multline}
    \left[\tilde h\, b^2-\tilde h_j b^2,\tilde h_i b^2\right]_{\expval{r_a,s_a}} \stackrel{b\to\infty}{\simeq}\\
    \prod_{\substack{0\leq m\leq r_a-1\\0\leq n\leq s_a-1}} \hspace*{-5pt} \left(\tilde{h}\!+\!\tilde{h}_{ij}-\frac{\mathrm{i}}{2} \sqrt{4 \tilde{h}_i\!-\!1} (2 m\!-\!r\!+\!1)-\frac{1}{4} (2 m\!+\!1) (2 m\!-\!2 r\!+\!1)-\frac14\right)b^2    \stackrel{b\to\infty}{\sim} b^{2r_as_a}\,,
    \label{eq:scaling_genPoch_heavy}
\end{multline}
and the case with $\tilde{h}+\tilde{h}_{ij}=0$
\begin{equation}
    \left[-\tilde{h}_i b^2,\tilde{h}_i b^2\right]_{\expval{r_a,s_a}} \stackrel{b\to\infty}{\sim}\begin{cases}
        b^{0} \quad &\text{if} \quad r_a=1\\
        b^{2r_a s_a} \quad &\text{if} \quad r_a>1
    \end{cases}\,.
\end{equation}
Note that we ignored the sums $\sum_{k=1}^{a-1}r_ks_k$ in the first argument of the generalized Pochhammers as they are sub-leading.

As one can expect from~\eqref{eq:hrs_scaling_b}, for most of the cases we consider, the main contribution arises from the $\expval{1,s}$ singular vectors.  For this reason, it is worth extracting the leading behavior of the generalized Pochhammers when $\expval{\boldsymbol{r},\boldsymbol{s}}=\expval{\boldsymbol{1},\boldsymbol{s}}\equiv\left\{ \expval{1,s_1},\dots,\expval{1,s_m}\right\}$ with $\sum_{a=1}^ms_a=\ell$.  In the case of light fields, the product of generalized Pochhammers simply reproduces the standard Pochhammer
\begin{equation}
		[h-h_j,h_i]_{\expval{\boldsymbol{1},\boldsymbol{s}}}\stackrel{b\to\infty}{=}(h+h_{ij})_\ell+\mathrm{O}\!\left(b^{-2}\right)\,,
        \label{eq:genPoch_1s_light}
\end{equation}
as one can verify from~\eqref{eq:scaling_genPoch_light}.  For heavy generic fields we have, from~\eqref{eq:scaling_genPoch_heavy},
\begin{equation}
    \left[\tilde h\, b^2-\tilde h_j b^2,\tilde h_i b^2\right]_{\expval{\boldsymbol{1},\boldsymbol{s}}} \stackrel{b\to\infty}{\simeq} b^{2\ell}\left(\tilde{h}+\tilde{h}_{ij}\right)^\ell\,.
    \label{eq:genPoch_1s_heavy}
\end{equation}
Having established the scaling of all factors in~\eqref{eq:block_coefficient_for_scaling}, we are ready to discuss various large-charge limits to leading order.


\subsubsection{The global limit}

The global limit corresponds to the case in which the central charge is taken to be large while all the conformal dimensions are light.  From the scaling analysis of~\eqref{eq:block_coefficient_for_scaling}, one can easily see that the scaling of generalized Pochhammers~\eqref{eq:scaling_genPoch_light} is always canceled by that of the corresponding inverse Zamolodchikov norm~\eqref{eq:qrs_scaling_b}.  What dictates which terms are leading are therefore the poles in~\eqref{eq:poles_scaling_b} and the combining factors in~\eqref{eq:combining_factor_scaling_b}, which can only contribute negative powers of $b$.  It is easy to see that to have the lowest power of $b$ in the combining factors~\eqref{eq:combining_factor_scaling_b} we must satisfy $r_1=\cdots=r_m$, while to have the lowest power of $b$ in the poles~\eqref{eq:poles_scaling_b} we must enforce $r_1=1$.  Combining the two, the leading order contributions emerge from $r_1=\cdots=r_m=1$ in the global limit.  Using~\eqref{eq:genPoch_1s_light}, we then have
\begin{equation}
		\mathcal{F}_h^{\boldsymbol{h}}(\cratio)\stackrel{b\to\infty}\simeq\sum_{\ell\ge 0} \left(h+h_{21}\right)_\ell \left(h+h_{34}\right)_\ell \left[\lim_{b\to\infty}\sum_{\boldsymbol{1}\cdot\boldsymbol{s}=\ell}\frac{q_{\expval{\boldsymbol{1},\boldsymbol{s}}}}{h-h_{\expval{1,s_1}}}\right]\cratio^{h+\ell}\,,
\end{equation}
and we can evaluate the object in the square brackets by realizing that
\begin{equation}
		\lim_{b\to\infty}\sum_{\boldsymbol{1}\cdot\boldsymbol{s}=\ell}\frac{q_{\expval{\boldsymbol{1},\boldsymbol{s}}}}{h-h_{\expval{1,s_1}}}=\frac{1}{(2h)_\ell}\left[\lim_{b\to\infty} \sum_{\boldsymbol{r}\cdot\boldsymbol{s}=\ell}\frac{q_{\expval{\boldsymbol{r},\boldsymbol{s}}}[h,h]_{\expval{\boldsymbol{r},\boldsymbol{s}}}}{h-h_{\expval{r_1,s_1}}}\right]=\frac{1}{(2h)_\ell\ell!}\,,
        \label{eq:global_limit_identity_OPE}
\end{equation}
which comes from the large-$b$ limit of~\eqref{eq:OPE_ell!identity}.  We are then left with
\begin{equation}
		\mathcal{F}_h^{\boldsymbol{h}}(\cratio)\stackrel{b\to\infty}{\simeq} \sum_{\ell\ge 0}\frac{\left(h+h_{21}\right)_\ell \left(h+h_{34}\right)_\ell}{(2h)_\ell\ell!}\cratio^{h+\ell}=\cratio^h \,{}_2F_1\!\left[\begin{array}{c} h+h_{21},h+h_{34}\\ 2h \end{array};\cratio\right]\,,
\end{equation}
which corresponds precisely to an $\mathfrak{sl}(2,\mathds{R})$ conformal block, as expected.


\subsubsection{Semi-classical limits}
Let us now consider the cases in which some of the conformal dimensions are taken to be heavy, that is $h_i \propto b^2$ and/or $h \propto b^2$. The results of this subsection are technically valid only for a small cross-ratio that scales either as $b^{-2}$ or $b^{-4}$, although we will not extract this scaling explicitly from $x$.

\paragraph{Case 1: generic heavy operators}

Let us consider first the case in which all the conformal dimensions, including all external and the exchanged operators, are heavy, that is $h_i,\,h \propto b^2$.  From the scaling of Pochhammers with heavy fields~\eqref{eq:scaling_genPoch_heavy} and the inverse Zamolochikov norms~\eqref{eq:qrs_scaling_b}, we have that the numerator in~\eqref{eq:block_coefficient_for_scaling} scales as
\begin{equation}
		\prod_{a=1}^m q_{\expval{r_a,s_a}}\left[\tilde{h} \,b^2-\tilde h_1 b^2,\tilde h_2 b^2\right]_{\expval{r_a,s_a}}\left[\tilde h\, b^2-\tilde h_4 b^2,\tilde h_3 b^2\right]_{\expval{r_a,s_a}} \stackrel{b\to\infty}{\sim} \prod_{a=1}^m b^{4s_a}\,,
        \label{eq:naive_scaling_heavy_semiclassical}
\end{equation}
which is independent on $r_a$ and thus has maximal power when $r_1=\cdots=r_m=1$.  Similarly to the global case, this is also the case in which the denominators have the lowest power.  We therefore expect the leading asymptotics of the conformal block to be produced just by the terms with $\expval{\boldsymbol{r},\boldsymbol{s}}=\expval{\boldsymbol{1},\boldsymbol{s}}$.

Using~\eqref{eq:genPoch_1s_heavy}, we get
\begin{equation}
		\mathcal{F}_{\tilde h b^2}^{(\tilde h_1 b^2,\tilde h_2 b^2,\tilde h_3 b^2,\tilde h_4 b^2)}(\cratio)\stackrel{b\to\infty}{\simeq}\sum_{\ell\ge 0} b^{4\ell}\left(\tilde h+\tilde h_{21}\right)^\ell \left(\tilde h+\tilde h_{34}\right)^\ell \left[\lim_{b\to\infty}\sum_{\boldsymbol{1}\cdot\boldsymbol{s}=\ell}\frac{q_{\expval{\boldsymbol{1},\boldsymbol{s}}}}{\tilde{h} \,b^2-h_{\expval{1,s_1}}}\right]\cratio^{h+\ell}\,.
        \label{eq:all_heavy_sc_identity_needed}
\end{equation}
The term in square brackets can be evaluated by taking the semi-classical limit of~\eqref{eq:OPE_ell!identity}
\begin{equation}
		\lim_{b\to\infty}\sum_{\boldsymbol{1}\cdot\boldsymbol{s}=\ell}\frac{q_{\expval{\boldsymbol{1},\boldsymbol{s}}}}{\tilde{h}\, b^2-h_{\expval{1,s_1}}}\simeq\lim_{b\to\infty}\frac{1}{(2\tilde{h} b^2)^{\ell}}\sum_{\boldsymbol{1}\cdot\boldsymbol{s}=\ell}\frac{q_{\expval{\boldsymbol{1},\boldsymbol{s}}}[\tilde{h}b^2,\tilde{h}b^2]_{\expval{\boldsymbol{1},\boldsymbol{s}}}}{\tilde h\, b^2-h_{\expval{1,s_1}}}\simeq\frac{1}{(2\tilde{h} b^2)^{\ell}\ell!}\,,
        \label{eq:OPEidentity_heavy_semiclassical}
\end{equation}
which can then be plugged in the expression above to obtain the exponentiated block
\begin{equation}
    \mathcal{F}_{\tilde h b^2}^{(\tilde h_1 b^2,\tilde h_2 b^2,\tilde h_3 b^2,\tilde h_4 b^2)}(\cratio)\stackrel{b\to\infty}{\simeq}\cratio^{b^2 \tilde{h}}\sum_{\ell\ge 0}\frac{\left(\tilde{h}+\tilde{h}_{21}\right)^\ell \left(\tilde{h}+\tilde{h}_{34}\right)^\ell}{(2\tilde{h})^{\ell}\ell!} b^{2\ell}\cratio^{\ell}=e^{b^{2}\tilde{h} \log \cratio+b^2\frac{\left(\tilde{h}+\tilde{h}_{21}\right) \left(\tilde{h}+\tilde{h}_{34}\right)}{2\tilde{h}} \cratio}\,.
    \label{eq:Semiclassical_block_all_heavy}
\end{equation}

The exponentiation of blocks in this semi-classical limit is well established in the literature~\cite{Zamolodchikov1986JETP,Zamolodchikov:1995aa,Besken:2019jyw}, where the function in the exponent proportional to $b^2$ is known to be related to the classical Liouville action~\cite{Zamolodchikov:1995aa} and is often named the \emph{classical conformal block}.  The procedure we followed here constitutes just a linear approximation to the classical block, as the full determination of it would require keeping track of all positive $b$-powers at every level.  It is however a fairly non-trivial check that our result perfectly agrees with e.g.~\cite[Eq.~8.10]{Zamolodchikov:1995aa} to leading order in~$\cratio$.  Moreover, as long as the exchanged operator is kept heavy $h\propto b^2$, one can recover the formula for the case where some or all the external operators are light just by sending the corresponding $\tilde{h}_i\to0$ in~\eqref{eq:Semiclassical_block_all_heavy}.

Finally, let us stress how the derivation of the leading term in the asymptotics performed in this paragraph may look convincing but is not completely rigorous in the way it was presented.  In fact, we studied which terms in the sum~\eqref{eq:Virasoro_blocks} had the largest scaling in $b$ and kept only the $\expval{\boldsymbol{1},\boldsymbol{s}}$ that naively scale as $b^{4\ell-2}$.  However, the application of the identity~\eqref{eq:OPEidentity_heavy_semiclassical} diminished their scaling to $b^{2\ell}$.  At this order, terms with different $\expval{\boldsymbol{r},\boldsymbol{s}}$, that scale between $b^{2\ell}$ and $b^{4\ell-2}$, could in principle also contribute. In Appendix~\ref{app:higher-r_subleading} we provide a more rigorous argument why $r>1$ contributions produce sub-leading terms in the limit considered here.


\paragraph{Case 2: special cases of heavy operators}

In the preceding semi-classical limit, we implicitly assumed that $\tilde{h}+\tilde{h}_{ij}\ne 0$ for both $ij=21$ and $ij=34$.  If this is not the case, the terms we focused on at every level are no longer present, and different terms need to be isolated.  While we will not prove these instances, from a level-by-level analysis of~\eqref{eq:Virasoro_blocks} we conjecture that the leading asymptotics of the semi-classical conformal blocks in this limit is
\begin{equation}
    \mathcal{F}_{\tilde h b^2}^{(\tilde h b^2+\tilde h_2 b^2,\tilde h_2 b^2,\tilde h_3 b^2,\tilde h_4 b^2)}(\cratio)\stackrel{b\to\infty}{\simeq}e^{b^2 \tilde{h} \log\cratio +b^2\frac{\tilde{h}_2 \left(\tilde{h}^2+2 \left(\tilde{h}_3+\tilde{h}_4\right) \tilde{h}-3 \left(\tilde{h}_3-\tilde{h}_4\right){}^2\right)}{4 \tilde{h} \left(4 \tilde{h}+3\right)}\cratio^2}\,,
    \label{eq:semiclassical_heavy_oneside_srule}
\end{equation}
when $\tilde{h}_1=\tilde h+\tilde{h}_2$, and
\begin{equation}
    \mathcal{F}_{\tilde h b^2}^{(\tilde h b^2+\tilde h_2 b^2,\tilde h_2 b^2,\tilde h_3 b^2,\tilde h b^2+\tilde h_3 b^2)}(\cratio)\stackrel{b\to\infty}{\simeq}e^{b^2 \tilde{h} \log\cratio + \frac{\tilde{h}_2 \tilde{h}_3}{4 \tilde{h}+3}b^2\cratio^2}\,,
        \label{eq:semiclassical_heavy_twoside_srule}
\end{equation}
when both $\tilde{h}_1=\tilde h+\tilde{h}_2$ and $\tilde{h}_4=\tilde h+\tilde{h}_3$.

On the one hand, note that in~\eqref{eq:semiclassical_heavy_oneside_srule} both $\tilde{h}_3$ and $\tilde{h}_4$ can be set to zero to recover cases with corresponding light external operators.  On the other hand, expression~\eqref{eq:semiclassical_heavy_twoside_srule} is instead the first case we encountered in which the internal conformal dimension can be taken to be light via the limit $\tilde{h}\to 0$.  This should also correspond to the asymptotics of the semiclassical vacuum block with all heavy external operators and with small cross-ratio $x \sim b^{-2}$. Note that the external dimensions, instead, cannot be taken to be light, so the known heavy-light limit of the vacuum block~\cite{Fitzpatrick:2015foa}
\begin{equation}
     \mathcal{F}_{0}^{(\tilde h_H b^2,\tilde h_H b^2,h_L,h_L)}(\cratio)\stackrel{b\to\infty}{\simeq}(1-z)^{h_L}\left(\frac{\sqrt{4\tilde{h}_H-1}}{2\sin\!\left( \sqrt{4\tilde{h}_H-1}\log\left(\sqrt{1-z}\right)\right)}\right)^{2h_L}\,,
     \label{eq:vacuum-heavy-light}
\end{equation}
cannot be reached from a limit of this expression.  Nevertheless, expression~\eqref{eq:vacuum-heavy-light} agrees level-by-level with the series expansion~\eqref{eq:Virasoro_blocks} in the heavy-light limit with $h=0$.

Finally, let us point out how both the numerators and denominators in the exponents of~\eqref{eq:semiclassical_heavy_oneside_srule} and~\eqref{eq:semiclassical_heavy_twoside_srule} are likely to originate from the singular vectors with $r=2$.  First, we have
\begin{equation}
    \left[-\tilde{h}_i b^2,\tilde{h}_i b^2\right]_{\expval{2,s}} \stackrel{b\to\infty}{\simeq}\prod_{n=0}^{s-1}\left[\frac{1}{2} b^2 \left(1- \mathrm i \sqrt{4 \tilde{h}_i-1}\right)\right]\left[\frac{1}{2} b^2 \left(1+ \mathrm i \sqrt{4 \tilde{h}_i-1}\right)\right]=b^{4s}\tilde{h}_i^s\,,
\end{equation}
which reproduce some of the factors in the numerators, and secondly
\begin{equation}
    \tilde{h}\,b^2-h_{\expval{2,s}}\stackrel{b\to\infty}{\simeq} \frac{b^2}{4} \left(4\tilde{h}+3\right)\,,
\end{equation}
which seem to originate from the $\expval{2,s}$ poles in the block expression~\eqref{eq:Virasoro_blocks}.


\paragraph{Case 3: generic heavy operators with light exchanged field}

If all of the external fields are taken to be heavy $h_i\propto b^2$, while the exchanged field is kept light $h=\mathrm{O}(b^0)$, the scaling analysis of the generalized Pochhammer symbols and inverse Zamolodchikov norms is described by~\eqref{eq:naive_scaling_heavy_semiclassical}.  We thus obtain an expression very similar to~\eqref{eq:all_heavy_sc_identity_needed} for the relevant terms in the series expansion
\begin{equation}
		\mathcal{F}_{h}^{(\tilde h_1 b^2,\tilde h_2 b^2,\tilde h_3 b^2,\tilde h_4 b^2)}(\cratio)\stackrel{b\to\infty}{\simeq}\sum_{\ell\ge 0} b^{4\ell}\tilde h_{21}^\ell \tilde h_{34}^\ell \left[\lim_{b\to\infty}\sum_{\boldsymbol{1}\cdot\boldsymbol{s}=\ell}\frac{q_{\expval{\boldsymbol{1},\boldsymbol{s}}}}{h-h_{\expval{1,s_1}}}\right]\cratio^{h+\ell}\,,
        \label{eq:heavy_lightexch_sc_identity_needed}
\end{equation}
where, importantly, the $h$ inside of the square brackets is not assumed to be large.  This means we can apply the same identity we had used in the global limit~\eqref{eq:global_limit_identity_OPE}. The conformal blocks in this limit can thus be resummed to a Bessel-Clifford $\mathcal{C}$ function
\begin{equation}
    \mathcal{F}_{h}^{(\tilde h_1 b^2,\tilde h_2 b^2,\tilde h_3 b^2,\tilde h_4 b^2)}(\cratio)\stackrel{b\to\infty}{\simeq}\cratio^{h}\sum_{\ell\ge 0} \frac{1}{\ell!}\frac{\tilde{h}_{21}^\ell \tilde{h}_{34}^\ell}{(2h)_{\ell}} b^{4\ell}\cratio^{\ell}=\Gamma(2h)\cratio^{h}\,\mathcal{C}_{2h-1}\!\left(b^4 \tilde h_{12}\tilde h_{34} \cratio\right)\,,
    \label{eq:semiclassical_BesselClifford}
\end{equation}
or, analogously, to a Bessel $J$ function
\begin{equation}
    \mathcal{F}_{h}^{(\tilde h_1 b^2,\tilde h_2 b^2,\tilde h_3 b^2,\tilde h_4 b^2)}(\cratio)\stackrel{b\to\infty}{\simeq}\Gamma (2 h) \left(b^2 \tilde{h}_{12}^{1/2} \tilde{h}_{34}^{1/2} \right)^{1-2 h} \sqrt\cratio \,J_{2 h-1}\!\left(2 b^2 \tilde{h}_{12}^{1/2} \tilde{h}_{34}^{1/2}\sqrt{\cratio }\right)\,.
\end{equation}
In this case, we can only set to zero at most one of the pair $\tilde{h}_1$, $\tilde{h}_2$ and at most one of the pair $\tilde{h}_3$, $\tilde{h}_4$ to obtain possible limits for light operators. Note that the formula presented here is valid in a regime where the cross-ratio $x\sim b^{-4}$ is smaller than the previously discussed cases.


\paragraph{Case 4: heavy-light limit with different external dimensions}

The last case we consider involves heavy fields on one side of the conformal block, $\tilde{h}_1\neq \tilde{h}_2\propto b^2$, and all other fields light $h$, $h_3$, $h_4=\mathrm{O}(b^0)$.  The overall scaling of the numerators of~\eqref{eq:block_coefficient_for_scaling}, in this case, is
\begin{equation}
		\prod_{a=1}^m q_{\expval{r_n,s_n}}\left[-\tilde h_1 b^2,\tilde h_2 b^2\right]_{\expval{r_n,s_n}}\left[h- h_4,h_3\right]_{\expval{r_n,s_n}} \stackrel{b\to\infty}{\sim} \prod_{a=1}^m b^{2s_a}\,,
        \label{eq:scaling_heavy-light_semiclassical}
\end{equation}
which is, once again, largest in all cases with $r_1=\cdots=r_m=1$.  Using both expressions~\eqref{eq:genPoch_1s_heavy} with $\tilde{h}=0$ and~\eqref{eq:genPoch_1s_light}, combined with the light-$h$ identity~\eqref{eq:global_limit_identity_OPE}, we obtain
\begin{equation}
    \mathcal{F}_h^{\boldsymbol{h}}(\cratio)\stackrel{b\to\infty}{=}\sum_{\ell\ge 0} \frac{1}{\ell!}\frac{\tilde{h}_{21}^\ell \left(h+h_{34}\right)_\ell}{(2h)^{\ell}} b^{2\ell}\cratio^{h+\ell}=\cratio^h\,{}_1F_1\left[\begin{array}{c}
       h+h_{34}   \\
          2h
    \end{array}; b^2 \tilde{h}_{21} \cratio \right]\,.
    \label{eq:semiclassical_1F1}
\end{equation}
This expression also covers the case in which one of the pair $\tilde{h}_1$, $\tilde{h}_2$ is sent to zero, with the corresponding field being therefore light.  This is the final case of semi-classical asymptotics we derive from our formula~\eqref{eq:Virasoro_blocks}. From these simple explorations it is clear that~\eqref{eq:Virasoro_blocks} has the potential to lead to novel expressions for Virasoro conformal blocks in special limits.


\section{Conclusions}

In this work, we used the inverse Shapovalov form and resolution of the identity obtained in~\cite{Fortin:2024xir} to determine explicit expressions for the Virasoro OPE~\eqref{eq:OPE} and for arbitrary four-point Virasoro conformal blocks on the Riemann sphere~\eqref{eq:Virasoro_blocks}.  Our main result for the blocks, \eqref{eq:Virasoro_blocks}, takes the form of a series expansion in the standard cross-ratio $\cratio=\frac{z_{12}z_{34}}{z_{13}z_{24}}$ with coefficients organized by the level of descendant fields and determined analytically.  The $\ell$-th coefficient of the series expansion~\eqref{eq:Virasoro_blocks} inherits from the resolution of the identity~\eqref{eq:resolution_identity} a finite combinatorial sum corresponding to all the ways in which level-$\ell$ descendants can be produced via composition of singular vector operators $L_{\expval{r,s}}$.  The same type of combinatorial sum also appears in the context of Zamolodchikov's recursion formula, as highlighted in Section~\ref{ssec:Zamolodchikov_comparison}.  Note that while our formula also shares the same explicit poles as Zamolodchikov's formula, our expression is fundamentally different from it, as one can easily notice by the presence of $h$ in the numerators.

In deriving~\eqref{eq:Virasoro_blocks}, we inserted the resolution of the identity~\eqref{eq:resolution_identity}, which depends on the action of the differential operators associated with singular vectors $\mathcal{L}_{\expval{r,s}}$, on Virasoro three-point functions.  This naively seems to depend on precise expressions for all Virasoro singular vectors, which are available in the literature but are very complicated~\cite{Benoit:1988aw,Bauer:1991ai,Millionshchikov2016,Watts:2024pnm}.  However, we have shown in Section~\ref{ssec:determine_OPE} how one can use the Virasoro fusion rules to circumvent the need for this knowledge, repackaging the action of the $\mathcal{L}_{\expval{r,s}}$ differential operators in terms of some multiplicative coefficient $\left[\alpha,\beta\right]_{\expval{r,s}}$ determined in~\eqref{eq:gen_Poch_explicit}.
This determines all coefficients of the series expansion~\eqref{eq:Virasoro_blocks}, which provides a new explicit way to work with Virasoro conformal blocks analytically.  Compared to other state-of-the-art expressions~\cite{Alday:2009aq,Alba:2010qc,Perlmutter:2015iya}, our formula strikes a balance between simplicity of the involved coefficients and an explicit representation-theoretic interpretation of its components.

In order to use our result for analytical applications, it is important to have good control over the non-standard finite sums present in the expression of the conformal blocks for arbitrary level $\ell$.  A step in this direction is given by the powerful identities we derived in Section~\ref{ssec:block_identity_limit} and Appendix~\ref{sec:app_OPE}; we expect these to be just the simplest, but more identities of this type are likely to be established.  Even with this reduced toolkit, we could analyze in Section~\ref{ssec:large_c} various large-charge limits of the Virasoro conformal blocks, resumming their expressions in various cases including the global limit and the asymptotics of some of the semiclassical limits.  We retrieve the well-known leading exponentiation for all heavy operators~\eqref{eq:Semiclassical_block_all_heavy}, but also discuss less explored cases: the asymptotic behaviors displayed in~\eqref{eq:semiclassical_heavy_oneside_srule}, \eqref{eq:semiclassical_heavy_twoside_srule}, \eqref{eq:semiclassical_BesselClifford}, and~\eqref{eq:semiclassical_1F1}---valid for small cross ratios---are, to our knowledge, new.  These small steps are indicative that an improved control over the finite sums in the blocks could bring insight for the construction of the complete classical block, corresponding to the leading exponent in the semiclassical limit.

Turning to numerical applications, the current state of the art is represented by the aforementioned $h$-recursion relation of Zamolodchikov, which produces a rapidly convergent expansion through the use of the nome variable $\mathfrak{q}$. This still remains the fastest approach to numerically evaluate conformal blocks with varying exchanged conformal weight $h$. However, our expression could be useful for applications that require to work with the standard cross-ratio $x$. Our expression has in fact, on the one hand, the same complexity of sums as Zamolodchikov's formula and, on the other hand, can be implemented to recycle all of the coefficients worked out at lower levels when computing higher levels. Given the natural appearance of the nome when considering conformal blocks in pillow geometry~\cite{Das:2017cnv}, it would be interesting to explore if the action of the differential operators $\mathcal{L}_{\expval{r,s}}$ could be expressed in a simple way also in that context, which may lead to an efficient series expansion in terms of the nome variable $\mathfrak{q}$.

From a more formal point of view, in light of the recent determination of a generalized Casimir operator $\mathcal{C}$ for the Virasoro algebra~\cite{Fortin:2024wcs}, it is interesting to highlight how the conformal blocks~\eqref{eq:Virasoro_blocks} must constitute concrete eigenfunctions for the infinite-order differential operators descending from $\mathcal{C}^{(i)}$ for any leg $i=1,\cdots,4$ and to that derived from $\mathcal{C}^{(12)}=\mathcal{C}^{(34)}$ (see Appendix~\ref{app:Casimir}).  We are not aware of other infinite-order eigenvalue equations of such complexity that have explicit solutions.

Finally, our new expression~\eqref{eq:Virasoro_blocks} may have the potential to shed new light on the general analytical structure of the Virasoro conformal blocks.  On the one hand, the analysis of the scaling of coefficients with the level may settle the debate on the radius of convergence of the Virasoro conformal block expansion in the cross-ratio~\cite[footnote~39]{Kusuki:2024gtq}.  On the other hand, as we show in Appendix~\ref{sec:app_DoubleGammaFunction}, the coefficients of the conformal blocks can be expressed in terms of some special combinations of the double Gamma function $\Gamma_b$, ubiquitous in the context of Liouville theory and with special properties under shifts in $b$ or $b^{-1}$.  This fact invites for thought on whether the Virasoro conformal blocks could belong to a class of special functions that naturally generalizes the hypergeometric functions but where the role of Gamma functions and Pochhammer symbols is replaced by their generalization on the lattice spanned by $b$ and~$b^{-1}$. Establishing the precise class of special functions the Virasoro conformal blocks belong to would have far-reaching consequences not only in CFT, but also in the mathematics of special functions, representation theory, and integrable systems.


\section*{Acknowledgements}

This work was supported by NSERC (JFF and LQ) and the US Department of Energy under grant DE-SC00-17660 (WS). LQ's research is partially supported
by the MUR PRIN contract 2022N9CTAE "Constraining strongly coupled quantum field theories using symmetry".

\appendix


\section{Virasoro OPE from the resolution of the identity}\label{sec:app_OPE}

In this appendix we demonstrate how to obtain the Virasoro OPE~\eqref{eq:OPE} from the resolution of the identity~\eqref{eq:resolution_identity}.  We also determine two interesting identities related to the identity limits for the three-point and four-point correlation functions, directly from the Virasoro OPE.


\subsection{Derivation of the Virasoro OPE}

Since the OPE is technically defined by its action on the vacuum, the first step is to act with the resolution of the identity on the product of primary fields $\phi_i(z)\phi_j(0)\ket{0}$ as in
\begin{equation}
\begin{aligned}
\phi_i(z)\phi_j(0)\ket{0}&=\sum_k\mathds{1}(c,h_k)\phi_i(z)\phi_j(0)\ket{0}\\
&=\sum_k\sum_{\substack{\ell\geq0\\\boldsymbol{r}\cdot\boldsymbol{s}=\ell}}L_{\expval{\boldsymbol{r},\boldsymbol{s}}}\ket{h_k}\!\bra{h_k}\frac{q_{\expval{\boldsymbol{r},\boldsymbol{s}}}}{L_0-h_{\expval{r_1,s_1}}}L_{\expval{\boldsymbol{r},\boldsymbol{s}}}^\dagger\phi_i(z)\phi_j(0)\ket{0}\\
&=\sum_k\bra{h_k}\phi_i(1)\ket{h_j}\sum_{\substack{\ell\geq0\\\boldsymbol{r}\cdot\boldsymbol{s}=\ell}}\frac{q_{\expval{\boldsymbol{r},\boldsymbol{s}}}}{h_k-h_{\expval{r_1,s_1}}}\frac{\bra{h_k}L_{\expval{\boldsymbol{r},\boldsymbol{s}}}^\dagger\phi_i(z)\ket{h_j}}{\bra{h_k}\phi_i(1)\ket{h_j}}L_{\expval{\boldsymbol{r},\boldsymbol{s}}}\phi_k(0)\ket{0}\,,
\end{aligned}
\end{equation}
with the help of the state-operator correspondence.  The match with the Virasoro OPE~\eqref{eq:OPE} is complete if
\begin{equation}
\frac{\bra{h_k}L_{\expval{\boldsymbol{r},\boldsymbol{s}}}^\dagger\phi_i(z)\ket{h_j}}{\bra{h_k}\phi_i(1)\ket{h_j}}=\frac{[h_k-h_j,h_i]_{\expval{\boldsymbol{r},\boldsymbol{s}}}}{z^{h_i+h_j-h_k-rs}}\,.
\label{eq:3ptLrs}
\end{equation}
To establish the identity~\eqref{eq:3ptLrs}, we use the action of the Virasoro generators on a primary field,
\begin{equation}
[L_n,\phi(z)]=-\mathcal{L}_n\phi(z)=[z^{n+1}\partial_z+h(n+1)z^n]\phi(z)\,,
\label{eq:Lnphi}
\end{equation}
and the identities
\begin{equation}
\mathcal{L}_{-\mu}z^a=[-a-h,h]_\mu z^{a-|\mu|}\,,\qquad(\mathcal{L}_{\mu})^Tz^a=(-1)^{l_\mu}[a+h,h]_\mu z^{a+|\mu|}\,,
\label{eq:action_of_differential_L_n_on_z}
\end{equation}
to compute two associated quantities,
\begin{equation}
\begin{gathered}
\frac{\bra{h_j}\phi_i(z)L_{\expval{\boldsymbol{r},\boldsymbol{s}}}\ket{h_k}}{\bra{h_j}\phi_i(1)\ket{h_k}}=\mathcal{L}_{\expval{\boldsymbol{r},\boldsymbol{s}}}\frac{\bra{h_j}\phi_i(z)\ket{h_k}}{\bra{h_j}\phi_i(1)\ket{h_k}}=\mathcal{L}_{\expval{\boldsymbol{r},\boldsymbol{s}}}\frac{1}{z^{h_i-h_j+h_k}}=\frac{[h_k-h_j,h_i]_{\expval{\boldsymbol{r},\boldsymbol{s}}}}{z^{h_i-h_j+h_k+rs}}\,,\\
\frac{\bra{h_k}L_{\expval{\boldsymbol{r},\boldsymbol{s}}}^\dagger\phi_i(z)\ket{h_j}}{\bra{h_k}\phi_i(1)\ket{h_j}}=\mathcal{L}_{\expval{\boldsymbol{r},\boldsymbol{s}}}^\dagger\frac{\bra{h_k}\phi_i(z)\ket{h_j}}{\bra{h_k}\phi_i(1)\ket{h_j}}=\mathcal{L}_{\expval{\boldsymbol{r},\boldsymbol{s}}}^\dagger\frac{1}{z^{h_i+h_j-h_k}}=\frac{[h_k-h_j,h_i]_{\expval{\boldsymbol{r},\boldsymbol{s}}}}{z^{h_i+h_j-h_k-rs}}\,,
\end{gathered}
\label{eq:3ptL}
\end{equation}
where
\begin{equation}
\mathcal{L}_{\expval{r,s}}\equiv v_{\expval{r,s}}^\mu\mathcal{L}_{-\mu}\,,\qquad\mathcal{L}_{\expval{r,s}}^\dagger\equiv(-1)^{l_\mu}v_{\expval{r,s}}^\mu(\mathcal{L}_\mu)^T\,,
\end{equation}
and
\begin{equation}
\mathcal{L}_{\expval{\boldsymbol{r},\boldsymbol{s}}}\equiv\mathcal{L}_{\expval{r_m,s_m}}\cdots\mathcal{L}_{\expval{r_1,s_1}}\,,\qquad\mathcal{L}_{\expval{\boldsymbol{r},\boldsymbol{s}}}^\dagger\equiv\mathcal{L}_{\expval{r_m,s_m}}^\dagger\cdots\mathcal{L}_{\expval{r_1,s_1}}^\dagger\,,
\end{equation}
following the notation in the main text.  The identity~\eqref{eq:3ptLrs} is thus verified with the help of the second equation in~\eqref{eq:3ptL}, which can be understood as the conjugate of the first equation in~\eqref{eq:3ptL}, both of which also appear in the computation of the four-point Virasoro blocks on the sphere.


\subsection{Identity limit of the Virasoro OPE}\label{app:IdentityLimitOPE}

From the Virasoro OPE~\eqref{eq:OPE}, it is straightforward to see that setting $\phi_i(z)=\mathds{1}$ with $h_i=0$ and enforcing the fusion rules $h_j=h_k=h$ lead to the consistent result $\phi(0)=\phi(0)$.  We can however produce an interesting set of identities from the Virasoro OPE~\eqref{eq:OPE} by choosing the other identity limit, namely $\phi_j(0)=\mathds{1}$ with $h_j=0$ and $h_i=h_k=h$ from the fusion rules.

Indeed, in this case we have
\begin{equation}
\phi(z)=\sum_{\substack{\ell\geq0\\\boldsymbol{r}\cdot\boldsymbol{s}=\ell}}\frac{q_{\expval{\boldsymbol{r},\boldsymbol{s}}}[h,h]_{\expval{\boldsymbol{r},\boldsymbol{s}}}}{h-h_{\expval{r_1,s_1}}}z^\ell L_{\expval{\boldsymbol{r},\boldsymbol{s}}}\phi(0)=\sum_{\substack{\ell\geq0\\\boldsymbol{r}\cdot\boldsymbol{s}=\ell}}\frac{q_{\expval{\boldsymbol{r},\boldsymbol{s}}}[h,h]_{\expval{\boldsymbol{r},\boldsymbol{s}}}}{h-h_{\expval{r_1,s_1}}}\Mdu{R}{\expval{\boldsymbol{r},\boldsymbol{s}}}{\lambda}z^\ell L_{-\lambda}\phi(0)\,,
\end{equation}
with
\begin{equation}
\Mdu{R}{\expval{\boldsymbol{r},\boldsymbol{s}}}{\lambda}=v_{\expval{r_1,s_1}}^{\mu_1}\cdots v_{\expval{r_m,s_m}}^{\mu_m}\Mdu{R}{\mu_2\mu_1}{\lambda_2}\Mdu{R}{\mu_3\lambda_2}{\lambda_3}\cdots\Mdu{R}{\mu_m\lambda_{m-1}}{\lambda}\,,\qquad L_{-\mu}L_{-\nu}=\Mdu{R}{\mu\nu}{\lambda}L_{-\lambda}\,,
\end{equation}
where $\Mdu{R}{\mu\nu}{\lambda}$ is simply an object encoding the reordering of the product of two sets of Virasoro generators $L_{-\mu}$ and $L_{-\nu}$ defined by two standard partitions $\mu$ and $\nu$ into a sum of Virasoro generators $L_{-\lambda}$ defined by standard partitions $\lambda$ (see~\cite{Fortin:2024wcs}).

Since the field at location $z$ is obtained by an action of only $L_1$ generators
\begin{equation}
\phi(z)=e^{zL_{-1}}\phi(0)=\sum_{\ell\geq0}\frac{1}{\ell!}(zL_{-1})^\ell\phi(0)\,,
\end{equation}
we conclude that
\begin{equation}
\sum_{\boldsymbol{r}\cdot\boldsymbol{s}=\ell}\frac{q_{\expval{\boldsymbol{r},\boldsymbol{s}}}[h,h]_{\expval{\boldsymbol{r},\boldsymbol{s}}}}{h-h_{\expval{r_1,s_1}}}\Mdu{R}{\expval{\boldsymbol{r},\boldsymbol{s}}}{\lambda}=\frac{1}{\ell!}\delta^{(1,\cdots,1)\lambda}\,,
\label{eq:OPE_Id_limit}
\end{equation}
which corresponds to $p(\ell)$ identities at level $\ell$.  Although the singular vector coefficients are complicated and the quantities $\Mdu{R}{\mu\nu}{\lambda}$ are tedious to compute, in the case $\lambda=(1,\cdots,1)$ we have $v_{\expval{r_a,s_a}}^{(1,\dots,1)}=\Mdu{R}{\expval{r_a,s_a}}{(1,\dots,1)}=1$ and~\eqref{eq:OPE_Id_limit} leads to the beautiful identity
\begin{equation*}
\sum_{\boldsymbol{r}\cdot\boldsymbol{s}=\ell}\frac{q_{\expval{\boldsymbol{r},\boldsymbol{s}}}[h,h]_{\expval{\boldsymbol{r},\boldsymbol{s}}}}{h-h_{\expval{r_1,s_1}}}=\frac{1}{\ell!}\,,
\end{equation*}
presented in~\eqref{eq:OPE_ell!identity} and which only depends on known quantities.


\subsection{Three-point correlation functions from the Virasoro OPE}\label{app:three-point_OPE}

Computing the three-point correlation function $\expval{\phi_k(w)\phi_i(z)\phi_j(0)}$ from the Virasoro OPE (or the resolution of the identity) is trivial when setting $w\to\infty$.  We can however generate an identity analogous to the identity limit of the four-point Virasoro blocks by keeping $w$ arbitrary.  From the Virasoro OPE~\eqref{eq:OPE}, we have
\begin{equation}
\begin{aligned}
\expval{\phi_k(w)\phi_i(z)\phi_j(0)}&=\sum_{k'}\frac{\bra{h_{k'}}\phi_i(1)\ket{h_j}}{z^{h_i+h_j-h_{k'}}}\sum_{\substack{\ell\geq0\\\boldsymbol{r}\cdot\boldsymbol{s}=\ell}}\frac{q_{\expval{\boldsymbol{r},\boldsymbol{s}}}[h_{k'}-h_j,h_i]_{\expval{\boldsymbol{r},\boldsymbol{s}}}}{h_{k'}-h_{\expval{r_1,s_1}}}z^\ell\bra{0}\phi_k(w)L_{\expval{\boldsymbol{r},\boldsymbol{s}}}\ket{h_{k'}}\\
&=\frac{\bra{h_k}\phi_i(1)\ket{h_j}}{z^{h_i+h_j-h_k}}\sum_{\substack{\ell\geq0\\\boldsymbol{r}\cdot\boldsymbol{s}=\ell}}\frac{q_{\expval{\boldsymbol{r},\boldsymbol{s}}}}{h_k-h_{\expval{r_1,s_1}}}[h_k-h_j,h_i]_{\expval{\boldsymbol{r},\boldsymbol{s}}}[h_k,h_k]_{\expval{\boldsymbol{r},\boldsymbol{s}}}\frac{z^\ell}{w^{2h_k+\ell}}\,,
\end{aligned}
\end{equation}
using the orthonormal Zamolodchikov metric and~\eqref{eq:3ptL}.

Comparing with the known result
\begin{equation}
\expval{\phi_k(w)\phi_i(z)\phi_j(0)}=\frac{\bra{h_k}\phi_i(1)\ket{h_j}}{(w-z)^{h_k+h_i-h_j}w^{h_k-h_i+h_j}z^{h_i+h_j-h_k}}\,,
\end{equation}
implies that
\begin{equation}
\sum_{\substack{\ell\geq0\\\boldsymbol{r}\cdot\boldsymbol{s}=\ell}}\frac{q_{\expval{\boldsymbol{r},\boldsymbol{s}}}}{h_k-h_{\expval{r_1,s_1}}}[h_k-h_j,h_i]_{\expval{\boldsymbol{r},\boldsymbol{s}}}[h_k,h_k]_{\expval{\boldsymbol{r},\boldsymbol{s}}}(z/w)^\ell=\frac{1}{(1-z/w)^{h_k+h_i-h_j}}\,,
\end{equation}
or
\begin{equation}
\sum_{\boldsymbol{r}\cdot\boldsymbol{s}=\ell}\frac{q_{\expval{\boldsymbol{r},\boldsymbol{s}}}}{h_k-h_{\expval{r_1,s_1}}}[h_k-h_j,h_i]_{\expval{\boldsymbol{r},\boldsymbol{s}}}[h_k,h_k]_{\expval{\boldsymbol{r},\boldsymbol{s}}}=\frac{(h_k-h_j+h_i)_\ell}{\ell!}\,.
\label{eq:3pt_Id_limit}
\end{equation}
The relation~\eqref{eq:3pt_Id_limit} is a refinement of~\eqref{eq:OPE_ell!identity} and is equivalent to the identity limit of four-point Virasoro blocks on the sphere~\eqref{eq:Virasoro_blocks} when $\phi_1=\mathds{1}$ or $\phi_4=\mathds{1}$, the identity limits for $\phi_2=\mathds{1}$ and $\phi_3=\mathds{1}$ being trivial.


\section{Block coefficients in terms of double Gamma functions}\label{sec:app_DoubleGammaFunction}

Since Barnes' double Gamma functions~\cite{BarnesDoubleGamma} are central objects in Liouville theory, it is natural that they appear in the Virasoro conformal blocks.  In this appendix, we re-express the generalized Pochhammer symbols~\eqref{eq:Pochhammerrs}, that are ubiquitous in our solution of the Virasoro blocks~\eqref{eq:Virasoro_blocks}, in terms of double Gamma functions.

The double Gamma function is defined by the functional equations~\cite{Eberhardt:2023mrq}
\begin{equation}
\Gamma_b(z+b)=\frac{\sqrt{2\pi}b^{bz-\frac{1}{2}}}{\Gamma(bz)}\Gamma_b(z)\,,\qquad\qquad\Gamma_b(z+b^{-1})=\frac{\sqrt{2\pi}b^{-b^{-1}z+\frac{1}{2}}}{\Gamma(b^{-1}z)}\Gamma_b(z)\,,
\label{eq:DoubleGammaFunction}
\end{equation}
up to an overall normalization which will ultimately cancel out in the generalized Pochhammer symbols.  For $b$ real and positive, the double Gamma function $\Gamma_b(z)$ has poles only at $z=-(mb+nb^{-1})$ for $m$ and $n$ non-negative integers, hence it can be thought of as an infinite product of the form
\begin{equation}
\Gamma_b(z)\propto\prod_{m,n\geq0}\frac{1}{(z+mb+nb^{-1})}\,.
\end{equation}

With that observation in mind, starting from~\eqref{eq:Pochhammerrs} written as (the order of one of the products has been changed)
\begin{equation}
[\alpha,\beta]_{\expval{r,s}}=\prod_{\pm}\prod_{\substack{0\leq m\leq r-1\\0\leq n\leq s-1}}\left(\sqrt{P_{\expval{r,s}}^2+\alpha}\pm\sqrt{P_{\expval{1,1}}^2-\beta}+P_{\expval{-r+1+2m,-s+1+2n}}\right)\,,
\label{eq:abprod}
\end{equation}
it suffices to introduce in the denominator two double Gamma functions that imply zeros for any $m\geq0,n\geq0$ and then eliminate exceeding zeros with four double Gamma functions in the numerator and two in the denominator to compensate for double counting.  This implies that
\begin{multline}
[\alpha,\beta]_{\expval{r,s}}\\=\prod_{\pm}\frac{\Gamma_b\!\left(\!\sqrt{\!P_{\expval{r,s}}^2\!+\!\alpha}\pm\!\sqrt{\!P_{\expval{1,1}}^2\!-\!\beta}+\!P_{\expval{r+1,-s+1}}\!\right)\!\Gamma_b\!\left(\!\sqrt{\!P_{\expval{r,s}}^2\!+\!\alpha}\pm\!\sqrt{\!P_{\expval{1,1}}^2\!-\!\beta}+\!P_{\expval{-r+1,s+1}}\!\right)}{\Gamma_b\!\left(\!\sqrt{\!P_{\expval{r,s}}^2\!+\!\alpha}\pm\!\sqrt{\!P_{\expval{1,1}}^2\!-\!\beta}+\!P_{\expval{-r+1,-s+1}}\!\right)\!\Gamma_b\!\left(\!\sqrt{\!P_{\expval{r,s}}^2\!+\!\alpha}\pm\!\sqrt{\!P_{\expval{1,1}}^2\!-\!\beta}+\!P_{\expval{r+1,s+1}}\!\right)}\,,
\end{multline}
for a total of four double Gamma functions in the numerator and in the denominator, canceling the overall normalization.

It is interesting to note that the generalized Pochhammer symbols~\eqref{eq:Pochhammerrs} can also be written in terms of a product of two ``double Pochhammer symbols'', generalized in a similar way to the double Gamma function, where integer shifts are replaced by shifts of integers times $b$ or $b^{-1}$ as in
\begin{equation}
(z)_{\expval{r,s}}\equiv\prod_{\substack{0\leq m\leq r-1\\0\leq n\leq s-1}}(z+mb+nb^{-1})\,.
\end{equation}
Starting again from~\eqref{eq:abprod}, the generalized Pochhammer symbols directly become
\begin{equation}
[\alpha,\beta]_{\expval{r,s}}=\prod_{\pm}\left(\sqrt{P_{\expval{r,s}}^2+\alpha}\pm\sqrt{P_{\expval{1,1}}^2-\beta}+P_{\expval{-r+1,-s+1}}\right)_{\!\expval{r,s}}\,.
\end{equation}
Likewise, the Zamolodchikov norms~\eqref{eq:qrs} can be rewritten in terms of double Pochhammer symbols as follows,
\begin{equation}
\frac{1}{q_{\expval{r,s}}}=\frac{2(-1)^{rs+1}}{rs}(b)_{\expval{r,s}}(-rb)_{\expval{r,s}}(b^{-1})_{\expval{r,s}}(-sb^{-1})_{\expval{r,s}}\,.
\end{equation}
%


\section{Proof that \texorpdfstring{$r>1$}{r>1} terms are sub-leading in the all-heavy semi-classical limit}\label{app:higher-r_subleading}

In Section~\ref{ssec:large_c}, we discussed conformal blocks at large central charge. The procedure relied on extracting the leading term in $b$ at every level and performing the sum over levels.  In the case where all operators are heavy and generic, which led to~\eqref{eq:Semiclassical_block_all_heavy}, we restricted our attention to  the $\expval{\boldsymbol{1},\boldsymbol{s}}$ terms. Even though this approach is valid, the argument we used in Section~\ref{ssec:large_c} was incomplete.  Below, we rectify this situation. 

In our analysis, we considered the ``naive scaling'' of each individual term $\expval{\boldsymbol{r},\boldsymbol{s}}$, for which the denominators enforce all $r_a$'s to be the same, and the numerators scale as
\begin{equation}
		\prod_{a=1}^m q_{\expval{r_a,s_a}}\left[\tilde{h} \,b^2-\tilde h_1 b^2,\tilde h_2 b^2\right]_{\expval{r_a,s_a}}\left[\tilde h\, b^2-\tilde h_4 b^2,\tilde h_3 b^2\right]_{\expval{r_a,s_a}} \stackrel{b\to\infty}{\sim} \prod_{a=1}^m b^{4s_a}\,.
        \tag{\ref{eq:naive_scaling_heavy_semiclassical}}
\end{equation}
However, we ultimately applied the identity
\begin{equation}
		\lim_{b\to\infty}\sum_{\boldsymbol{1}\cdot\boldsymbol{s}=\ell}\frac{q_{\expval{\boldsymbol{1},\boldsymbol{s}}}}{\tilde{h}\, b^2-h_{\expval{1,s_1}}}\simeq\lim_{b\to\infty}\frac{1}{(2\tilde{h} b^2)^{\ell}}\sum_{\boldsymbol{1}\cdot\boldsymbol{s}=\ell}\frac{q_{\expval{\boldsymbol{1},\boldsymbol{s}}}[\tilde{h}b^2,\tilde{h}b^2]_{\expval{\boldsymbol{1},\boldsymbol{s}}}}{\tilde h\, b^2-h_{\expval{1,s_1}}}\simeq\frac{1}{(2\tilde{h} b^2)^{\ell}\ell!}\,,
        \tag{\ref{eq:OPEidentity_heavy_semiclassical}}
\end{equation}
which lowered the order of the sum over $\expval{\boldsymbol{1},\boldsymbol{s}}$ terms to $b^{2\ell}$.  To argue that these terms are leading, we have to show that contributions with at least one $r_a>1$ lead to scaling lower than $b^{2\ell}$.

In order to proceed with our proof, let us first define a useful quantity:
\begin{equation}
    \mathcal{S}=\sum_{\expval{r,s}\in\{\expval{r_1,s_1},\dots,\expval{r_m,s_m}\}}\hspace{-1.4cm}s\delta_{r,1}\,,
\end{equation}
which counts how much of the level-$\ell$ singular vector decomposition $\{\expval{r_1,s_1},\dots,\expval{r_m,s_m}\}$ is made of singular vectors with $r=1$.  We similarly define
\begin{equation}
    \mathcal{T}=\sum_{\expval{r,s}\in\{\expval{r_1,s_1},\dots,\expval{r_m,s_m}\}}\hspace{-1.4cm}rs(1-\delta_{r,1})\,,
\end{equation}
that counts how much of the level is made of singular vectors with $r>1$.  Trivially, $\mathcal{T}=\ell-\mathcal{S}$.  The sequence $\expval{\boldsymbol{r},\boldsymbol{s}}=\{\expval{r_1,s_1},\dots,\expval{r_m,s_m}\}$ can be decomposed into maximal clusters of adjacent $\expval{1,s_a}$ singular vectors, which we dub clusters of $\mathcal{S}$ type, separated by clusters of $\expval{r_{a'},s_{a'}}$ singular vectors with all the $r_{a'}>1$, of $\mathcal{T}$ type.  For the $i$-th cluster we can then define both an $\mathcal{S}_i$ and a $\mathcal{T}_i$, only one of which can be nonzero according to the type of the cluster. 

This decomposition into clusters is useful to prove the auxiliary result
\begin{equation}
    \sum_{\boldsymbol{1}\cdot\boldsymbol{s}=\ell}q_{\expval{\boldsymbol{1},\boldsymbol{s}}}\stackrel{b\to\infty}{=}\mathrm{O}\!\left(b^{-2\ell}\right) \qquad \text{if}\quad \ell>1\,,
    \label{eq:sum_q1s_scalin1g_b}
\end{equation}
for which we use induction.  Using~\eqref{eq:sum_of_qrs=0} and~\eqref{eq:qrs_scaling_b}, we can prove for $\ell=2$
\begin{equation}
    q_{\expval{1,1}\expval{1,1}}+q_{\expval{1,2}}=-q_{\expval{2,1}}=\mathrm{O}\!\left(b^{-4}\right)\,.
\end{equation}
Let us now assume that~\eqref{eq:sum_q1s_scalin1g_b} holds for level $\ell-1$.  Using~\eqref{eq:sum_of_qrs=0}, we can rewrite the sum $\sum_{\expval{\boldsymbol{1},\boldsymbol{s}}=\ell}q_{\expval{\boldsymbol{1},\boldsymbol{s}}}$ as a sum over $\expval{\boldsymbol{r},\boldsymbol{s}}$ made of $k_{\mathcal{S}}$ clusters of type $\mathcal{S}$ and $k_{\mathcal{T}}\ge 1$ clusters of $\mathcal{T}$ type
\begin{equation}
\begin{aligned}
    \sum_{\expval{\boldsymbol{1},\boldsymbol{s}}=\ell}q_{\expval{\boldsymbol{1},\boldsymbol{s}}}&=-\sum_{\boldsymbol{r}\cdot\boldsymbol{s}=\ell}q_{\expval{\boldsymbol{r},\boldsymbol{s}}}\\
    &\text{with} \quad \boldsymbol{r}=
        (\underbrace{r_1,\dots,r_{\mathcal{T}_1}}_{\mathcal{T}_1},\underbrace{1,\dots,1}_{\mathcal{S}_1},r'_1,\dots ) \quad \text{or}\quad \boldsymbol{r}=
        (\underbrace{1,\dots,1}_{\mathcal{S}_1},\underbrace{r_1,\dots,r_{\mathcal{T}_1}}_{\mathcal{T}_1},1,\dots )\,.
        \end{aligned}
        \label{eq:decomposition_clusters}
\end{equation}
More specifically, if we consider e.g. the first possibility of the two above for $\boldsymbol{r}$, we have
\begin{equation}
    -\sum_{\boldsymbol{r}\cdot\boldsymbol{s}=\ell}q^{(\mathcal{T}_1)}_{\expval{\boldsymbol{r},\boldsymbol{s}}}\frac{1}{h_{\expval{r_{\mathcal{T}_1},s_{\mathcal{T}_1}}}+r_{\mathcal{T}_1}s_{\mathcal{T}_1}-h_{\expval{1,s_{\mathcal{T}_1+1}}}}q_{\expval{\boldsymbol{1},\boldsymbol{s}}}^{(\mathcal{S}_1)}\cdots\stackrel{b\to\infty}{\simeq}-\sum_{\boldsymbol{r}\cdot\boldsymbol{s}=\ell}q^{(\mathcal{T}_1)}_{\expval{\boldsymbol{r},\boldsymbol{s}}}\frac{4}{1-r_{\mathcal{T}_1}^2}\frac{1}{b^2}q_{\expval{\boldsymbol{1},\boldsymbol{s}}}^{(\mathcal{S}_1)}\cdots\,,
    \label{eq:clusters_separated_by_b^2}
\end{equation}
where we highlighted that the combining factors~\eqref{eq:combining_factor_scaling_b} that connect a $\mathcal{T}$ cluster with an $\mathcal{S}$ cluster are independent on the structure of the latter, and always introduce a factor of $b^{-2}$.  The same happens when connecting an $\mathcal{S}$ cluster with a $\mathcal{T}$ cluster.  With this in mind, we can directly see from~\eqref{eq:qrs_scaling_b} that each individual $q^{(\mathcal{T}_i)}_{\expval{\boldsymbol{r},\boldsymbol{s}}}$ scales at most as
\begin{equation}
    b^{\frac{4\mathcal{T}_i}{\min(r_1,\dots,r_m)}}\,,
    \label{eq:boundfromabove_num_scaling}
\end{equation}
which is strictly lower than $b^{-2\mathcal{T}_i}$.  If we instead consider the sum over each $\mathcal{S}_i$ cluster, by inductive assumption it scales at most as $b^{-2\mathcal{S}_i}$ if $\mathcal{S}_i\ge2$, while it trivially scales as $b^0$ if $\mathcal{S}_i=1$.  Keeping in mind that, since $\mathcal{S}<\ell$, each $\mathcal{S}$ cluster is always accompanied by a factor of $b^{-2}$ as in~\eqref{eq:clusters_separated_by_b^2}, we can include this factor with that of the $\mathcal{S}_i$ cluster to argue that the sum over the cluster scales at most as $b^{-2\mathcal{S}_i}$ for every $\mathcal{S}_i\ge1$.  Combining the scalings coming from all clusters, we now have that
\begin{equation}
    \sum_{\expval{\boldsymbol{1},\boldsymbol{s}}=\ell}q_{\expval{\boldsymbol{1},\boldsymbol{s}}} \stackrel{b\to\infty}{=} \mathrm{O}\!\left(b^{-2\sum_{i=1}^{k_{\mathcal{T}}}\mathcal{T}_i-2\sum_{j=1}^{k_{\mathcal{S}}}\mathcal{S}_j}\right)=\mathrm{O}\!\left(b^{-2\ell}\right)\,,
\end{equation}
proving the auxiliary result~\eqref{eq:sum_q1s_scalin1g_b}.

We can now finally proceed with the proof that any level-$\ell$ term that has an $r_a\ne 1$ in the all-heavy semiclassical limit scales slower than $b^{-2\ell}$, and thus does not contribute to~\eqref{eq:Semiclassical_block_all_heavy} at leading order in $b$.

We will divide the proof in two steps:
\begin{enumerate}
    \item Show that all terms with $\mathcal{S}=0$ individually have a scaling lower than $b^{2\ell}$;
    \item Show that, when $0<\mathcal{S}<\ell$, the sum over all the $\expval{1,s_a}$ that make up the sub-level $\mathcal{S}$ reduces the overall scaling to a value below~$b^{2\ell}$.
\end{enumerate}
To prove the first point, let us recall that the scaling of individual terms is made of the product of the numerator scaling~\eqref{eq:naive_scaling_heavy_semiclassical}, the scaling of the poles that always contribute a $b^{-2}$, and the scaling of the combining factors~\eqref{eq:combining_factor_scaling_b} which depends on how many times adjacent $r_a$'s differ.  For the factors in the numerator~\eqref{eq:naive_scaling_heavy_semiclassical}, the scaling is bounded by above by~\eqref{eq:boundfromabove_num_scaling}.  Taking into account the naive scaling of the $h$ pole which reduces the scaling by a factor $b^{-2}$, it is immediately apparent that if $\mathcal{S}=0$, we have the upper bound
\begin{equation}
    b^{2\ell-2}\,.
\end{equation}
This clearly shows that terms with $\mathcal{S}=0$ are sub-leading compared to the $\mathcal{S}=\ell$ terms we focused on in the main text, which have scaling $b^{2\ell}$.

We now know that the only way the scaling could be above $b^{2\ell}$ is to have at least $\mathcal{S}\ge 1$.  Since the case $\mathcal{S}=\ell$ leads to the terms of order $b^{2\ell}$ we considered in the main text, here we will focus on $\mathcal{S}<\ell$.  In this case, since the terms that do not contribute to $\mathcal{S}$ have $r_a\ge 2$, we can refine the upper bound~\eqref{eq:boundfromabove_num_scaling} to
\begin{equation}
    b^{2\left(\ell-\mathcal{S}\right)+4\mathcal{S}}=b^{2\ell+2\mathcal{S}}\,.
    \label{eq:temporary_scaling_b_S-clusters}
\end{equation}

At this point, let us consider a decomposition in $\mathcal{S}$ and $\mathcal{T}$ clusters akin to~\eqref{eq:decomposition_clusters}, but including all factors of the block coefficients, not just the $q_{\expval{\boldsymbol{r},\boldsymbol{s}}}$.  The generalized Pochhammers associated with $\mathcal{S}$ clusters only depend on their level $\mathcal{S}_i$
\begin{equation}
		[h-h_j,h_i]_{\expval{\boldsymbol{1},\boldsymbol{s}}}\stackrel{b\to\infty}{=}(h+h_{ij})_{\mathcal{S}_i}+\mathrm{O}\!\left(b^{-2}\right)\,.
\end{equation}
Similarly to the proof of the auxiliary result, the combining factors between clusters are independent on the structure of the $\mathcal{S}$ cluster and introduce a $b^{-2}$ factor.  Finally, if $\expval{\boldsymbol{r},\boldsymbol{s}}$ starts with an $\mathcal{S}$ cluster, the pole term is at leading order independent of $s_1$:
\begin{equation}
    \frac{1}{\tilde h b^2-h_{\expval{1,s_1}}}\stackrel{b\to\infty}{\simeq} \frac{1}{\tilde{h} b^2}\,.
\end{equation}
This means that the sums over the $\mathcal{S}$ clusters are all of the form discussed in the auxiliary result, implying that each $\mathcal{S}$-cluster of level $\mathcal{S}_i$ introduces a scaling of at least $b^{-2\mathcal{S}_i}$.  Performing the sum over all $\mathcal{S}$ clusters, we then have that these introduce an overall $b^{-2\mathcal{S}}$ factor, which when combined with~\eqref{eq:temporary_scaling_b_S-clusters} and the $b^{-2}$ of the pole, results in the upper bound for the scaling:
\begin{equation}
    b^{2\ell +2\mathcal{S}-2\mathcal{S}-2}=b^{2\ell-2}\,.
\end{equation}
This concludes the proof that all contributions to~\eqref{eq:Semiclassical_block_all_heavy} coming from $\expval{\boldsymbol{r},\boldsymbol{s}}$ with at least one $r_a\ne1$ are sub-leading compared to the $\expval{\boldsymbol{1},\boldsymbol{s}}$ terms we discussed in the main text.


\section{Virasoro Casimirs and blocks}\label{app:Casimir}

This appendix obtains the Virasoro Casimir differential operators for the four-point Virasoro conformal blocks on the sphere and demonstrates that the latter are eigenfunctions of the former with the appropriate eigenvalues.


\subsection{Virasoro Casimir differential operators}

Let Latin letters typeset in roman [$\iu=(i_1,\cdots,i_{l_\iu})$, $\ju=(j_1,\cdots,j_{l_\ju})$, \textit{etc.}] represent ordered partitions.  Then, for the ordered partition $\iu$, the parts satisfy $1\leq i_1,i_2,\cdots,i_{l_\iu}$ where $l_\iu$ is the length of $\iu$ (or the number of parts in $\iu$), $|\iu|=\sum_{k=1}^{l_\iu}i_k$ is the level of $\iu$ (or the sum of its parts), and by analogy to standard partitions
\begin{equation}
L_{-\iu}\ket{c,h}\equiv L_{-i_{l_\iu}}\cdots L_{-i_1}\ket{c,h}\,,
\end{equation}
even though the ordered partitions form an overcomplete basis.  Following~\cite{Fortin:2024wcs}, the Casimirs of the Virasoro algebra $\mathcal{C}$, which depend on the central charge element $\hat{c}$ and the Virasoro generators $L_{n\in\mathbb{Z}}$ through functions $M_0(\hat{c},L_0)$ that encode their eigenvalue, are given by
\begin{equation}
\begin{aligned}
\mathcal{C}(M_0)&=\sum_{\ell\geq0}\sum_{\substack{\iu\\|\iu|=\ell}}(-1)^{l_\iu}\sum_{\bs{r}^j\cdot\bs{s}^j=i_j}\left[\prod_{j=l_\iu}^1L_{\expval{\bs{r}^j,\bs{s}^j}}\right]\{M_0(\hat{c},L_0+\ell)-M_0(\hat{c},L_0+\xi_{l_\iu})\}\\
&\qquad\times\left[\prod_{j=l_\iu}^1\frac{q_{\expval{\bs{r}^j,\bs{s}^j}}}{L_0+\xi_j-h_{\expval{r_1^j,s_1^j}}}\right]\left[\prod_{j=1}^{l_\iu}L_{\expval{\bs{r}^j,\bs{s}^j}}^\dagger\right]\,,
\label{eq:Cas}
\end{aligned}
\end{equation}
where it is understood that $\prod_{j=l_\iu}^1a_j\equiv a_{l_\iu}\cdots a_1$, contrary to $\prod_{j=1}^{l_\iu}a_j\equiv a_1\cdots a_{l_\iu}$, that the $\ell=0$ contribution is $M_0(\hat{c},L_0)$ and that
\begin{equation}
\xi_j=\sum_{1\leq k<j}i_k,\qquad\qquad\xi_1\equiv0,
\end{equation}
for the ordered partition $\iu$.

Extracting the conformally-covariant prefactor
\begin{equation}
\Omega=\frac{1}{z_{12}^{h_1+h_2}z_{34}^{h_3+h_4}}\left(\frac{z_{24}}{z_{14}}\right)^{h_{12}}\left(\frac{z_{14}}{z_{13}}\right)^{h_{34}},
\label{eq:Omega}
\end{equation}
from the four-point correlation functions, the four-point Virasoro blocks~\eqref{eq:Virasoro_blocks} are expressible as
\begin{equation}
\mathcal{F}_h^{\bs{h}}(x)=\Omega^{-1}\frac{\bra{0}\phi_1(z_1)\phi_2(z_2)\mathds{1}(c,h)\phi_{3}(z_{3})\phi_4(z_4)\ket{0}}{\bra{h_1}\phi_2(1)\ket{h}\!\bra{h}\phi_3(1)\ket{h_4}}\,.
\label{eq:VirB}
\end{equation}
Since the Virasoro Casimirs~\eqref{eq:Cas} must act diagonally on the projection of the correlator $\bra{0}\phi_1(z_1)\phi_2(z_2)\mathds{1}(c,h)\phi_{3}(z_{3})\phi_4(z_4)\ket{0}$, one must have
\begin{equation}
\bra{0}\!\phi_1(z_1)\phi_2(z_2)\mathds{1}(c,h)\mathcal{C}(M_0)\phi_{3}(z_{3})\phi_4(z_4)\!\ket{0}=M_0(c,h)\!\bra{0}\!\phi_1(z_1)\phi_2(z_2)\mathds{1}(c,h)\phi_{3}(z_{3})\phi_4(z_4)\!\ket{0}.
\end{equation}
In the $z_1\to\infty$, $z_4\to0$ limit, this identity becomes
\begin{equation}
\bra{h_1}\phi_2(z_2)\mathds{1}(c,h)\mathcal{C}(M_0)\phi_{3}(z_{3})\ket{h_4}=M_0(c,h)\bra{h_1}\phi_2(z_2)\mathds{1}(c,h)\phi_{3}(z_{3})\ket{h_4}\,.
\end{equation}
Considering that $\mathcal{C}\supset\left[\prod_{j=l_\iu}^1L_{\expval{\bs{r}^j,\bs{s}^j}}\right]L_0^p\left[\prod_{j=1}^{l_\iu}L_{\expval{\bs{r}^j,\bs{s}^j}}^\dagger\right]$, $L_n$ commutes with $\mathds{1}(c,h)$ and that $L_0\ket{h_4}=h_4\ket{h_4}$ while $\bra{h_1}L_{n<0}=L_{n>0}\ket{h_4}=0$, the action of the Virasoro Casimirs can be transformed into differential operators $\mathcal{D}(M_0)$ such that
\begin{equation}
\widehat{\Omega}\mathcal{D}(M_0)\widehat{\Omega}^{-1}\bra{h_1}\phi_2(z_2)\mathds{1}(c,h)\phi_{3}(z_{3})\ket{h_4}=M_0(c,h)\bra{h_1}\phi_2(z_2)\mathds{1}(c,h)\phi_{3}(z_{3})\ket{h_4},
\end{equation}
where
\begin{equation}
\widehat{\Omega}=\lim_{\substack{z_1\to\infty\\z_4\to0}}z_1^{2h_1}\Omega=\frac{z_2^{h_{12}}}{z_3^{h_3+h_4}}\,.
\end{equation}
In terms of the four-point Virasoro blocks~\eqref{eq:VirB}, this means that
\begin{equation}
\mathcal{D}(M_0)\mathcal{F}_h^{\bs{h}}(x)=M_0(c,h)\mathcal{F}_h^{\bs{h}}(x)\,,
\end{equation}
where
\begin{equation}
\begin{aligned}
\mathcal{D}(M_0)&=\lim_{y\to1}\frac{x^{h_3+h_4}}{y^{h_3+h_4+h_{21}}}\sum_{\ell\geq0}\sum_{\substack{\iu\\|\iu|=\ell}}(-1)^{l_\iu}\sum_{\bs{r}^j\cdot\bs{s}^j=i_j}\left[\prod_{j=l_\iu}^1\mathcal{L}_{\expval{\bs{r}^j,\bs{s}^j}}\right]\left[\prod_{j=l_\iu}^1\mathcal{L}_{\expval{\bs{r}^j,\bs{s}^j}}^\dagger\right]\\
&\qquad\times\left[\prod_{j=l_\iu}^1\frac{q_{\expval{\bs{r}^j,\bs{s}^j}}}{-\mathcal{L}_0+h_4+\xi_j-h_{\expval{r_1^j,s_1^j}}}\right]\\
&\qquad\times\{M_0(c,-\mathcal{L}_0+h_4+\ell)-M_0(c,-\mathcal{L}_0+h_4+\xi_{l_\iu})\}\frac{y^{h_3+h_4+h_{21}}}{x^{h_3+h_4}}\,,
\end{aligned}
\label{eq:expression_Casimir_as_limit}
\end{equation}
and $x=z_3/z_2$, $y=1/z_2$.  Here both $\mathcal{L}_0$ and $\mathcal{L}_{\expval{\bs{r},\bs{s}}}^\dagger$ act with $z_3=x/y$ and $h_3$ while $\mathcal{L}_{\expval{\bs{r},\bs{s}}}$ acts with $z_2=1/y$ and $h_2$.  Thus, we can use the identities
\begin{equation}
\begin{gathered}
\frac{x^{h_3+h_4}}{y^{h_3+h_4+h_{21}}}(-\mathcal{L}_0+h_4)\frac{y^{h_3+h_4+h_{21}}}{x^{h_3+h_4}}=\vartheta\,,\\
\frac{x^{h_3+h_4}}{y^{h_3+h_4+h_{21}}}(-\mathcal{L}_{n>0})\frac{y^{h_3+h_4+h_{21}}}{x^{h_3+h_4}}=(x/y)^n(\vartheta+h_3n-h_4)\,,\\
\frac{x^{h_3+h_4}}{y^{h_3+h_4+h_{21}}}(\mathcal{L}_{n<0})\frac{y^{h_3+h_4+h_{21}}}{x^{h_3+h_4}}=y^{-n}[\vartheta+h_2(-n)-h_1+y\partial_y]\,,\\
\end{gathered}
\end{equation}
where $\vartheta=x\partial_x$, to evaluate all actions of generators in the expression~\eqref{eq:expression_Casimir_as_limit}. Keeping track of all powers of $x/y$ and $y$ introduced by the $\mathcal{L}_n$, and using~\eqref{eq:action_of_differential_L_n_on_z} together with~\eqref{eq:abrs},
we can express the action of the Casimir in terms of generalized Pochhammer symbols
\begin{equation}
\begin{aligned}
\mathcal{D}(M_0)&=\sum_{\ell\geq0}x^\ell\sum_{\substack{\iu\\|\iu|=\ell}}(-1)^{l_\iu}\{M_0(c,\vartheta+\ell)-M_0(c,\vartheta+\xi_{l_\iu})\}\\
&\qquad\times\sum_{\bs{r}^j\cdot\bs{s}^j=i_j}\left[\prod_{j=1}^{l_\iu}\frac{q_{\expval{\bs{r}^j,\bs{s}^j}}}{\vartheta+\xi_j-h_{\expval{r_1^j,s_1^j}}}[\vartheta+\xi_j-h_1,h_2]_{\expval{\bs{r}^j,\bs{s}^j}}[\vartheta+\xi_j-h_4,h_3]_{\expval{\bs{r}^j,\bs{s}^j}}\right]\,,
\end{aligned}
\label{eq:VirCasDiffOps}
\end{equation}
where the $\ell=0$ term is understood to be $M_0(c,\vartheta)$ and the order of the products is now unimportant.

Clearly, the Virasoro Casimir differential operators~\eqref{eq:VirCasDiffOps} are infinite-order differential operators that must have eigenfunctions given by the four-point Virasoro blocks
\begin{equation}
\begin{aligned}
\mathcal{F}_h^{\bs{h}}(x)&=\sum_{\substack{\ell\geq0\\\bs{r}\cdot\bs{s}=\ell}}\frac{q_{\expval{\bs{r},\bs{s}}}}{h-h_{\expval{r_1,s_1}}}[h-h_1,h_2]_{\expval{\bs{r},\bs{s}}}[h-h_4,h_3]_{\expval{\bs{r},\bs{s}}}x^{h+\ell}\,,
\label{eq:VirBlocks}
\end{aligned}
\end{equation}
as will now be shown.


\subsection{Virasoro Casimir differential operators and four-point Virasoro blocks}

Acting on a power of the cross-ratio $x$, the Virasoro Casimir differential operators~\eqref{eq:VirCasDiffOps} give
\begin{equation}
\begin{aligned}
\mathcal{D}(M_0)x^{h+n}=&\sum_{\ell\geq0}x^{h+n+\ell}\sum_{\substack{\iu\\|\iu|=\ell}}(-1)^{l_\iu}\{M_0(c,h+n+\ell)-M_0(c,h+n+\xi_{l_\iu})\}\\
&\times\sum_{\bs{r}^j\cdot\bs{s}^j=i_j}\left[\prod_{j=1}^{l_\iu}\frac{q_{\expval{\bs{r}^j,\bs{s}^j}}[h+n+\xi_j-h_1,h_2]_{\expval{\bs{r}^j,\bs{s}^j}}[h+n+\xi_j-h_4,h_3]_{\expval{\bs{r}^j,\bs{s}^j}}}{h+n+\xi_j-h_{\expval{r_1^j,s_1^j}}}\right]\,.
\end{aligned}
\end{equation}
Thus, for the four-point Virasoro blocks~\eqref{eq:VirBlocks} to satisfy $\mathcal{D}(M_0)\mathcal{F}_h^{\bs{h}}(x)=M_0(c,h)\mathcal{F}_h^{\bs{h}}(x)$, this implies
\begin{equation}
\begin{aligned}
&M_0(c,h)\sum_{\bs{r}\cdot\bs{s}=\ell}\frac{q_{\expval{\bs{r},\bs{s}}}}{h-h_{\expval{r_1,s_1}}}[h-h_1,h_2]_{\expval{\bs{r},\bs{s}}}[h-h_4,h_3]_{\expval{\bs{r},\bs{s}}}\\[6pt]
&=\sum_{0\leq n\leq\ell}\sum_{\substack{\iu\\|\iu|=\ell-n}}(-1)^{l_\iu}\{M_0(c,h+\ell)-M_0(c,h+n+\xi_{l_\iu})\}\\[-6pt]
&\qquad\quad\times\sum_{\bs{r}^j\cdot\bs{s}^j=i_j}\left[\prod_{j=1}^{l_\iu}\frac{q_{\expval{\bs{r}^j,\bs{s}^j}}[h+n+\xi_j-h_1,h_2]_{\expval{\bs{r}^j,\bs{s}^j}}[h+n+\xi_j-h_4,h_3]_{\expval{\bs{r}^j,\bs{s}^j}}}{h+n+\xi_j-h_{\expval{r_1^j,s_1^j}}}\right]\\[6pt]
&\qquad\quad\times\sum_{\bs{r}\cdot\bs{s}=n}\frac{q_{\expval{\bs{r},\bs{s}}}}{h-h_{\expval{r_1,s_1}}}[h-h_1,h_2]_{\expval{\bs{r},\bs{s}}}[h-h_4,h_3]_{\expval{\bs{r},\bs{s}}}\,,
\end{aligned}
\end{equation}
which is the Casimir constraint that must be verified.  Separating the $n=0$ and $n=\ell$ terms, the Casimir constraint becomes
\begin{equation}
\begin{aligned}
\{M_0&(c,h)-M_0(c,h+\ell)\}\sum_{\bs{r}\cdot\bs{s}=\ell}\frac{q_{\expval{\bs{r},\bs{s}}}}{h-h_{\expval{r_1,s_1}}}[h-h_1,h_2]_{\expval{\bs{r},\bs{s}}}[h-h_4,h_3]_{\expval{\bs{r},\bs{s}}}\\
=&\sum_{\substack{\iu\\|\iu|=\ell}}(-1)^{l_\iu}\{M_0(c,h+\ell)-M_0(c,h+\xi_{l_\iu})\}\\[-14pt]
&\qquad\quad\times\sum_{\bs{r}^j\cdot\bs{s}^j=i_j}\left[\prod_{j=1}^{l_\iu}\frac{q_{\expval{\bs{r}^j,\bs{s}^j}}[h+\xi_j-h_1,h_2]_{\expval{\bs{r}^j,\bs{s}^j}}[h+\xi_j-h_4,h_3]_{\expval{\bs{r}^j,\bs{s}^j}}}{h+\xi_j-h_{\expval{r_1^j,s_1^j}}}\right]\\[4pt]
&+\sum_{1\leq n\leq\ell-1}\sum_{\substack{\iu\\|\iu|=\ell-n}}(-1)^{l_\iu}\{M_0(c,h+\ell)-M_0(c,h+n+\xi_{l_\iu})\}\\[-6pt]
&\qquad\quad\times\sum_{\bs{r}^j\cdot\bs{s}^j=i_j}\left[\prod_{j=1}^{l_\iu}\frac{q_{\expval{\bs{r}^j,\bs{s}^j}}[h+n+\xi_j-h_1,h_2]_{\expval{\bs{r}^j,\bs{s}^j}}[h+n+\xi_j-h_4,h_3]_{\expval{\bs{r}^j,\bs{s}^j}}}{h+n+\xi_j-h_{\expval{r_1^j,s_1^j}}}\right]\\[4pt]
&\qquad\quad\times\sum_{\bs{r}\cdot\bs{s}=n}\frac{q_{\expval{\bs{r},\bs{s}}}}{h-h_{\expval{r_1,s_1}}}[h-h_1,h_2]_{\expval{\bs{r},\bs{s}}}[h-h_4,h_3]_{\expval{\bs{r},\bs{s}}}\,.
\end{aligned}
\end{equation}
Merging the $1\leq n\leq\ell-1$ into the ordered partition $\iu$ gives
\begin{equation}
\begin{aligned}
\{M_0&(c,h)-M_0(c,h+\ell)\}\sum_{\bs{r}\cdot\bs{s}=\ell}\frac{q_{\expval{\bs{r},\bs{s}}}}{h-h_{\expval{r_1,s_1}}}[h-h_1,h_2]_{\expval{\bs{r},\bs{s}}}[h-h_4,h_3]_{\expval{\bs{r},\bs{s}}}\\
=&\sum_{\substack{\iu\\|\iu|=\ell}}(-1)^{l_\iu}\{M_0(c,h+\ell)-M_0(c,h+\xi_{l_\iu})\}\\[-12pt]
&\qquad\times\sum_{\bs{r}^j\cdot\bs{s}^j=i_j}\left[\prod_{j=1}^{l_\iu}\frac{q_{\expval{\bs{r}^j,\bs{s}^j}}[h+\xi_j-h_1,h_2]_{\expval{\bs{r}^j,\bs{s}^j}}[h+\xi_j-h_4,h_3]_{\expval{\bs{r}^j,\bs{s}^j}}}{h+\xi_j-h_{\expval{r_1^j,s_1^j}}}\right]\\[6pt]
&+\sum_{\substack{\iu\\|\iu|=\ell\\l_\iu\geq2}}(-1)^{l_\iu-1}\{M_0(c,h+\ell)-M_0(c,h+\xi_{l_\iu})\}\\[-12pt]
&\qquad\times\sum_{\bs{r}^j\cdot\bs{s}^j=i_j}\left[\prod_{j=1}^{l_\iu}\frac{q_{\expval{\bs{r}^j,\bs{s}^j}}[h+\xi_j-h_1,h_2]_{\expval{\bs{r}^j,\bs{s}^j}}[h+\xi_j-h_4,h_3]_{\expval{\bs{r}^j,\bs{s}^j}}}{h+\xi_j-h_{\expval{r_1^j,s_1^j}}}\right]\,.
\end{aligned}
\end{equation}
Separating the first sum over ordered partitions $\iu$ into a sum with $l_\iu=1$ and a sum with $l_\iu\geq2$, the RHS becomes identical to the LHS and proves the Casimir constraint, showing that the four-point Virasoro conformal blocks are eigenfunctions of the Virasoro Casimir differential operators.

We note that although any choice of $M_0(\hat{c},L_0)$ is allowed in the Virasoro Casimirs, it is not easy to find a non-trivial function $M_0(\hat{c},L_0)$ that has the same eigenvalues for submodules in degenerate Verma modules interpreted as highest-weight modules~\cite{FeiginFuchs:Casimir}.




\end{document}